\documentclass[a4paper,11pt]{article}
\pdfoutput=1 

\usepackage{jheppub} 
\usepackage{tikz}
\usepackage{tikz-cd}

\newcommand{\bea}{\begin{eqnarray}}
\newcommand{\eea}{\end{eqnarray}}

\newcommand{\nn}{\nonumber}

\usepackage[utf8]{inputenc}
\usepackage{hyperref}
\usepackage{amsfonts}
\usepackage{amsmath}
\usepackage{amssymb}
\usepackage{mathrsfs}  
\usepackage{color}
\usepackage{physics}
\usepackage{multicol}
\usepackage{tikz}

\usepackage{amsmath,bbm,array,amsfonts,graphicx,wrapfig,lscape,float,mathtools,multirow,longtable,amsthm}
\usepackage{stackrel}
\usepackage[all]{xy}
\usepackage{caption}
\usepackage{subcaption}
\usepackage{multirow}
\usepackage[bottom]{footmisc}
\usepackage{mathrsfs}
\usepackage{tikz}
\usepackage{bm}
\usepackage{color}
\usepackage{enumerate}
\usetikzlibrary{decorations.pathreplacing}
\usepackage{diagbox}
\numberwithin{equation}{section}
\numberwithin{figure}{section}
\numberwithin{table}{section}
\usepackage{pgfplots}
\pgfplotsset{compat=1.14}
\usepackage{makecell}
\usepackage{lscape}
\usepackage[utf8]{inputenc}
\usepackage{mathtools}
\usepackage{todonotes}
\usepackage{epigraph}

\usepackage[autostyle=true]{csquotes}
\newtheorem{definition}{Definition}[section]
\newtheorem{theorem}{Theorem}[section]
\newtheorem{corollary}{Corollary}[theorem]
\newtheorem{lemma}[theorem]{Lemma}
\newtheorem{proposition}[theorem]{Proposition}
\newtheorem{conjecture}[theorem]{Conjecture} 
\newtheorem{remark}{Remark}
\newtheorem{example}{Example}

\usepackage[toc,page]{appendix}
\usepackage[inline]{enumitem}

\newcommand{\commentout}[1]{}

\title{Reflexions on Mahler: \\
\qquad Dessins, Modularity and Gauge Theories}

\author[a,b]{Jiakang Bao,}
\author[b,a,c,d]{Yang-Hui He,}
\author[e,b]{Ali Zahabi}

	\affiliation[a]{
		Department of Mathematics, City, University of London, EC1V 0HB, UK}
	\affiliation[b]{
	    London Institute for Mathematical Sciences, Royal Institution, London W1S 4BS, UK}
	\affiliation[c]{
		Merton College, University of Oxford, OX1 4JD, UK}
	\affiliation[d]{
		School of Physics, NanKai University, Tianjin, 300071, P.R. China}

    \affiliation[e]{Institut de Math\'ematiques de Bourgogne, Universit\'e Bourgogne Franche-Comt\'e, France}

	\emailAdd{jiakang.bao@city.ac.uk}
	\emailAdd{hey@maths.ox.ac.uk}
    \emailAdd{zahabi.ali@gmail.com}

\preprint{
		\begin{flushright}
			LIMS-2021-016
		\end{flushright}
	}

\abstract{We provide a unified framework of Mahler measure, dessins d'enfants, and gauge theory. With certain physically motivated Newton polynomials from reflexive polygons, the Mahler measure and the dessin are in one-to-one correspondence. From the Mahler measure, one can construct a Hauptmodul for a congruence subgroup of the modular group, which contains the subgroup associated to the dessin. We also discuss their connections to the quantum periods of del Pezzo surfaces, as well as certain elliptic pencils. In brane tilings and quiver gauge theories, the modular Mahler flow might shed light on the inequivalence amongst the three different complex structures $\tau_{R,G,B}$. We also study how, in F-theory, 7-branes and their monodromies arise in the context of dessins.
}

\begin{document} 
\maketitle

\epigraph{I am the child of artists and have always lived among artists and, also, I'm one myself.}{Gustav Mahler}

\section{Introduction and Summary}\label{intro}
Mahler measure has been an important concept in number theory and geometry over the last several decades \cite{mahler1962some,boyd1980speculations,boyd1981kronecker,villegas1999modular,boyd2002mahler,bertin2013mahler}. While its definition is straight-forward and requires nothing beyond complex analysis - it is the geometric mean of the log-absolute value of a Laurent polynomial over the unit $n$-torus - it has profound consequences in a multitude of mathematical disciplines. Moreover, a generalization of the Mahler measure to an arbitrary torus, called the Ronkin function, has been a useful tool in the study of tropical geometry and is closely related to amoebae of the polynomial.

Perhaps more surprisingly, the Mahler measure has found applications in physics as well. For instance, its connection to volume conjecture and knot invariants \cite{boyd2002mahler2} has appeared in the context of topological strings \cite{Dijkgraaf:2009sb}. In dimer models, Mahler measures and Ronkin functions reveal the generating function of the perfect matchings \cite{Kenyon:2003uj}. As a result, they can be used to count BPS states/Donaldson-Thomas invariants using crystal melting and quiver quantum mechanics, which also give rise to emergent Calabi-Yau (CY)\footnote{Strictly speaking, this should be called Gorenstein as it is singular (and non-compact). Nevertheless, we shall abuse nomenclature, and call them CY, as in most of the literature.} geometry \cite{Ooguri:2009ri}.

Recently, Mahler measure has been more deeply explored in quiver gauge theories \cite{Stienstra:2005wz,Zahabi:2018znq,Zahabi:2019kdm,Zahabi:2020hwu,Bao:2021fjd}, which has been proposed as a ``universal measure'' for toric quivers. In particular, Mahler measure has monotonic behaviour along the so-called {\it Mahler flow} between different ``limits'' of the quiver gauge theory. Along the Mahler flow, it encodes a phase transition that has nice interpretations in both mathematics and physics. Furthermore, the Mahler measure has also been found to have intimate relation with $a$-maximization, and also enjoys salient properties under dualities \cite{Bao:2021fjd}.

In parallel, dessins d'enfants (children's drawings) \cite{girondo2012introduction} have constituted a core object in algebraic geometry and number theory since Grothendieck's Esquisse d'un Programme (sketch of a programme) \cite{Grothendieck1984sketch}. Following Belyi's theorem \cite{belyui1980galois}, these bipartite graphs can be nicely connected to algebraic curves. The study of dessins has then been led to the vast areas of Galois theory, modularity, and, more recently, congruence subgroups and monstrous moonshine \cite{He:2012jn,He:2013eqa,He:2015vua,Tatitscheff:2018aht,He:2020eva}. Due to the relationship to polynomial equations defining Riemann surfaces, dessins appear in the study of Seiberg-Witten theory, as points in the Coulomb branch \cite{Ashok:2006br}. They have then applied to various aspects in physics, including both $\mathcal{N}=1$ and $\mathcal{N}=2$ quivers, conformal blocks etc \cite{Stienstra:2007dy,Jejjala:2010vb,Bose:2014lea,Bao:2021vxt}.

Now, it was demonstrated in \cite{villegas1999modular} that the Mahler measure has certain expansion whose building blocks behave as modular forms. Therefore, it would be natural to expect some deep connection between Mahler measure and dessins due to the emergence of modularity. In particular, so-called reflexive polygons provide a nice playground since their Newton polynomials define elliptic curves.

\paragraph{Main Results:}
In this paper, we shall focus on modular Mahler measures for reflexive polygons.
The family of elliptic curves defined by the Newton polynomials with parameter $k$ furnishes the Klein's $j$-invariant as a meromorphic function $j(k) : \mathbb{P}^1 \to \mathbb{P}^1$. Of particular interest here would be the so-called tempered families that put certain restrictions on the coefficients of the Newton polynomials. We find that a subset of these families, which we call {\it maximally tempered}, give a one-to-one correspondence between Mahler measures and dessins.

A priori, there does not seem to be anything special for the maximally tempered coefficients except that they are non-zero binomial numbers along each edge of a reflexive polygon. However, it has a salient interpretation in physics. When constructing quiver theories from brane tilings, each lattice point in the Newton polygon is associated with some perfect matchings/gauged linear sigma model fields \cite{Franco:2005rj,Franco:2005sm,Feng:2005gw}. The maximally tempered coefficients are exactly the numbers of perfect matchings for the lattice points.

We find that the dessins obtained in such way are invariant under specular duality (Proposition \ref{samedessin}). On the other hand, how Mahler measures behave under specular duality has been studied in \cite{Bao:2021fjd}. Here, with the special maximally tempered coefficients, we find that Mahler measures are invariant for specular duals (Proposition \ref{samemahler} and Corollary \ref{samemahlerref}). Thus, the one-to-one correspondence between Mahler measures and dessins are automatic (Remark \ref{mahlerdessin}). As is known, specular duality preserves the master space of the gauge theory. However, different toric phases are often not related by such a duality\footnote{Yet, they have the same Mahler measure and dessin as the Newton polynomial does not change.}. Therefore, the Mahler measure and the dessin should encode some information of the master space.

To be more precise, the maximally tempered cases, despite their physical meaning in the brane tilings, are not always that well-behaved in the sense of relating the Mahler measure and the dessins as detailed in Proposition \ref{samedessin}. On the other hand, for some \emph{minimally tempered} cases for the reflexive polygons, they also show some nice properties in the connections to other mathematical objects. Nevertheless, for all the maximally tempered cases and two minimally tempered cases, they correspond to del Pezzo surfaces via the quantum periods of these Fano varieties (Table \ref{GX} and Equations \eqref{GX1}, \eqref{GX2}).

Calculations of the modular Mahler measure show that certain modular quantities are related to some congruence subgroups via their Hauptmoduln. In fact, they contain the congruence subgroups associated to the dessin (Conjecture \ref{hauptmodulconj1} and \ref{hauptmodulconj2}).
Besides, as the Mahler measure is derived from several modular forms (with singularities), one can naturally apply the results in \cite{kontsevich2001periods} and study the Mahler measure in terms of $j$-invariants (Proposition \ref{mjperiodalgebraic}, Equations \eqref{mjDE} and \eqref{mjtrop}).

Besides the two sets of congruence subgroups as mentioned above, the reflexive polygons are also related to certain elliptic pencils, giving rise to another set of congruence subgroups that are associated to these elliptic pencils. We may then compare all these subgroups (Propositions \ref{Beauvilleprop} and \ref{ellpenprop2}). In this context, some minimally tempered cases also enjoy such properties while some maximally tempered ones are not that well-behaved. This is in fact in line with the aforementioned connections between the Mahler measure and the dessins. We also summarize this in Table \ref{summarizecompare} below.

For quiver gauge theories, there has been a long-standing puzzle about the relationship between (i) $\tau_R$, the complex structure of the torus in the isoradial $a$-maximized brane tiling, (ii) $\tau_B$, the complex structure from the dessin, and (iii) $\tau_G$, the complex structure from the torus fibration in the mirror \cite{Jejjala:2010vb,Hanany:2011bs,He:2012xw}. It turns out that they do not necessarily coincide. In erudite terms, $\tau_{R,G,B}$ are different points along the Mahler flow. We also conjecture part of the relations among the complex structures (Conjecture \ref{taubtaurcondition}).

Modularity also has various important applications in many other topics such as (rational) conformal fied theories, black holes etc. Here, we show that the dessins give the positions of 7-branes in F-theory compactifications (Proposition \ref{branefaceblack}), and the monodromies of the 7-branes are related to the monodromy (cartographic) groups of the dessins.

Another interesting topic is the BPS states of non-critical strings in F-theory. In particular, the Gromov-Witten (GW) invariants from zero-size instantons have been computed in \cite{Lerche:1996ni,Klemm:1996hh}. It was later observed in \cite{Stienstra:2005wy} (also remarked in \cite{Stienstra:2005ns}) that modular expansions in Mahler measure recover the GW invariants from the instanton expansions of Yukawa coupling (with certain 4-cycle vanishing in the embedding CY geometry). Here, we give a dictionary between the quantities on the Mahler and GW sides (Table \ref{dictionary}).

Overall, the Mahler measure and the dessins are ``unified'' in the sense that they are in one-to-one correspondence with each other and that they both arise in the context of quiver gauge theories. In particular, the Mahler measure is determined by some (maximally) tempered Laurent polynomials in this paper. This leaves one free parameter $k$ which parametrizes the ``moduli space'' of an elliptic curve. The parameter $k$ can in turn be expressed as an expansion of the complex structure of a torus, which implements the concept of modular Mahler measure first introduced in \cite{villegas1999modular}. On the other hand, the dessins obtained from these Laurent polynomials are naturally associated to some congruence subgroups of SL$(2,\mathbb{Z})$. It turns out that these subgroups can have close relations with the subgroups whose Hauptmoduln are given by $k$ on the Mahler measure side. As mentioned above, the discussions on modularity would also appear in physics such as different complex structures in brane tilings and quiver gauge theories, the enumerations of Gromov-Witten invariants etc.

\paragraph{Organization}
The paper is organized as follows. In \S\ref{background}, we briefly review all the required background in mathematics and physics. 
Then in \S\ref{modularity}, we discuss the modular Mahler measure and its relation to dessins. In \S\ref{Fthy}, we venture to some connections to F-theory and 7-branes. In Appendix \ref{Pzw}, we list all the Newton polynomials considered in this paper, with coefficients dubbed maximally or minimally tempered. In Appendix \ref{minimally}, we have some discussions on the minimally tempered cases, similar to those on the maximally tempered ones in the main context.
In Appendix \ref{mj}, we mention some connections of the Mahler measure to the $j$-invariants and periods in the sense of Kontsevich and Zagier. We discuss the different complex structures of $\mathbb{T}^2$ arised in the context of brane tilings in Appendix \ref{taubtaur}. In Appendix \ref{GW}, we mention some relations between modular Mahler measure and GW invariants.

Since the Mahler measure and the dessins for the reflexive polygons are related to various objects in the paper. We summarize the main differences in these connections in Table \ref{summarizecompare}.
\begin{table}[h]
\centering
\begin{tabular}{c|c|c|c|c|c}
 & $\substack{\text{No.5$\sim$10,}\\ \text{13, 15, 16 (max)}}$ & $\substack{\text{No.1$\sim$4}\\ \text{(max)}}$ & $\substack{\text{No.11, 12, 14}\\ \text{(max)}}$ & $\substack{\text{No.7, 13}\\ \text{(min)}}$ & $\substack{\text{No.1, 4}\\ \text{(min)}}$ \\ \hline
$\substack{j(k)/1728\\ \text{Belyi?}}$ & $\checkmark$ & $\checkmark$ & $\substack{\text{require further maps}\\ \text{on the parameter}}$ & $\times$ & $\checkmark$ \\ \hline
dessins? & index 12 & index $<12$ & $\substack{\text{require further maps}\\ \text{on the parameter}}$ & $\times$ & index 12 \\ \hline
$\substack{\text{corresponding to}\\ \text{del Pezzo surfaces?}}$ & $\checkmark$ & $\checkmark$ & $\checkmark$ & $\checkmark$ & $\times$ \\ \hline
$\substack{\text{corresponding to}\\ \text{elliptic pencils?}}$ & Beauville & others & $\times$ & $\times$ & Beauville \\ \hline
$\substack{\text{congruence}\\ \text{subgroups}}$ & $\Gamma_p\leq\Gamma_d\leq\Gamma_k$ & $\Gamma_p\leq\Gamma_d\stackrel{?}{\leq}\Gamma_k$ & N/A & N/A & $\Gamma_p\leq\Gamma_d\leq\Gamma_k$
\end{tabular}
\caption{The numbers labelling the reflexive polygons can be found in Figure \ref{refpolygons}. The brackets with ``max'' and ``min'' indicate the coefficients associated to the Newton polynomials which will be explained below. The subscripts of the congruence subgroups $\Gamma$ are understood as follows. The ones associated to the elliptic pencils are labelled with $p$ while the ones from the dessins have label $d$. For $k$, this stands for the free parameter $k$ in the Newton polynomial in the context of modular Mahler measure. In this paper, the index of a congruence subgroup always means the index in the group PSL$(2,\mathbb{Z})$. The detailed statements can be found in the following sections.}\label{summarizecompare}
\end{table}

\section{Dramatis Personae}\label{background}
In this section, we give a brief review on the main topics of this study, namely the Mahler measure, reflexive polygons, elliptic curves, dessins etc.

\subsection{Mahler Measure}\label{mahler}
As introduced in \cite{mahler1962some}, the (logarithmic) \textbf{Mahler measure} is\footnote{In much of the literature, the Mahler measure is defined to be $\exp(\mathtt{m}(P))$, but we will exclusively work with this logarithmic version \eqref{def:mahler}.}
\begin{equation}\label{def:mahler}
    \begin{split}
        \mathtt{m}(P):&=\int_0^1\dots\int_0^1\log|P(\exp(2\pi i\theta_1),\dots,\exp(2\pi i\theta_n))|\textup{d}\theta_1\dots\textup{d}\theta_n\\
        &=\frac{1}{(2\pi i)^n}\int_{|z_1|=1}\dots\int_{|z_n|=1}\log|P(z_1,\dots,z_n)|\frac{\textup{d}z_1}{z_1}\dots\frac{\textup{d}z_n}{z_n}
    \end{split}
\end{equation}
for a non-zero Laurent polynomial $P(\bm{z})=P(z_1,\dots,z_n)\in\mathbb{C}[z_1^{\pm1},\dots,z_n^{\pm1}]$. Properties of Mahler measure have been extensively studied. For instance, it is invariant under $\bm{z}\rightarrow\pm\bm{z}^M$ for any non-zero matrix $(M_{ij})_{n\times n}\in\text{GL}(n,\mathbb{Z})$, where $\bm{z}^M=\left(\prod\limits_{i=1}^nz_i^{M_{i1}},\dots,\prod\limits_{i=1}^nz_i^{M_{in}}\right)$.

In this paper, we will focus on two-variable Laurent polynomials, viz, $\bm{z}=(z,w)$. Moreover, the Laurent polynomials are always of the form 
\begin{equation}\label{kNP}
P(z,w)=k-p(z,w) \ ,
\end{equation}
where $p(z,w)$ does not have a constant term. We further require that $|k|>\max\limits_{|z|=|w|=1}|p(z,w)|$. Therefore, \eqref{def:mahler} becomes
\begin{equation}
        \mathtt{m}(P)
        =
        \text{Re}\left(\frac{1}{(2\pi i)^2}\int_{|z|=|w|=1}\log(k-p(z,w))\frac{\textup{d}z}{z}\frac{\textup{d}w}{w}\right).
\end{equation}
For convenience, let us define
\begin{equation}
    m(P):=\frac{1}{(2\pi i)^2}\int_{|z|=|w|=1}\log(P(z,w))\frac{\textup{d}z}{z}\frac{\textup{d}w}{w}
\end{equation}
such that $\mathtt{m}(P)=\text{Re}(m(P))$ \footnote{In general, if we have real $k$ satisfying $k>\max\limits_{|z|=|w|=1}|p(z,w)|$, then $\mathtt{m}(P)=m(P)$.}.

Then, we may use the formula series expansion of $\log(k-p(z,w))$ in $p$, which converges uniformly on the support of the integration path, arriving at
\begin{equation}
    m(P)=\log k-\int_0^{k^{-1}}(u_0(t)-1)\frac{\text{d}t}{t},
\end{equation}
where
\begin{equation}
    u_0(k)=\frac{1}{(2\pi i)^2}\int_{|z|=|w|=1}\frac{1}{1-k^{-1}p(z,w)}\frac{\textup{d}z}{z}\frac{\textup{d}w}{w}.
\end{equation}
Since we will be mainly dealing with $m(P)$, we shall henceforth call $m(P)$ the Mahler measure as well for brevity.

It is not hard to see that
\begin{equation}
    \frac{\text{d}m(P)}{\textup{d}\log k}=k\frac{\text{d}m(P)}{\text{d}k}=u_0(k).
\end{equation}
Physically, this is called the \textbf{Mahler flow equation} \cite{Bao:2021fjd}. It reveals the monotonic behaviour of the Mahler measure\footnote{It is also conjectured that this could be extended to $|k|=0$ although we are not considering this extended region here. In such case, we shall take the Mahler flow equation as the definition for $u_0$.}:
\begin{proposition}
The Mahler measure $\mathtt{m}(P)$ strictly increases when $|k|$ increases from $\max\limits_{|z|=|w|=1}|p(z,w)|$ to $\infty$.
\end{proposition}

In terms of $\lambda=k^{-1}$, the Mahler flow equation reads
\begin{equation}
    \frac{\text{d}m}{\textup{d}\log\lambda}=\lambda\frac{\text{d}m}{\text{d}\lambda}=-u_0(\lambda).\label{mahlerfloweqn}
\end{equation}

\begin{example}
Let us consider $P=k-z-z^{-1}-w-w^{-1}$. The quantities $m$ and $u_0$ are given as
\begin{equation}
    m(P)=-\log(\lambda)-2\lambda^2{}_4F_3\left(1,1,\frac{3}{2},\frac{3}{2};2,2,2;16\lambda^2\right),\quad u_0(\lambda)={}_2F_1\left(\frac{1}{2},\frac{1}{2};1;16\lambda^2\right).\label{mu0ex}
\end{equation}
For detailed steps, the readers are referred to, for example, the appendix in \cite{Bao:2021fjd}. They have the expansions
\begin{align}
    &m(P)=-\log(\lambda)-2\lambda^2-9\lambda^4-\frac{200}{3}\lambda^6-\frac{1225}{2}\lambda^8-\frac{31752}{5}\lambda^{10}-\dots\\
    &u_0(k)=1+4\lambda^2+36\lambda^4+400\lambda^6+4900\lambda^8+63504\lambda^{10}+\dots
\end{align}
around $\lambda=0$ (or equivalently, $k=\infty$). More examples can be found in Table \ref{refmahler} along with \eqref{mandu0} below. We plot the Mahler flow as
\begin{equation}
    \includegraphics[width=7cm]{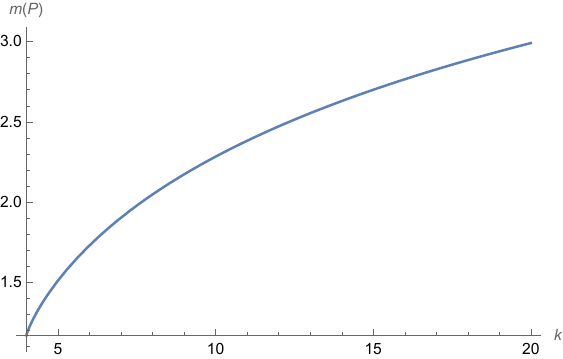},\qquad
    \includegraphics[width=7cm]{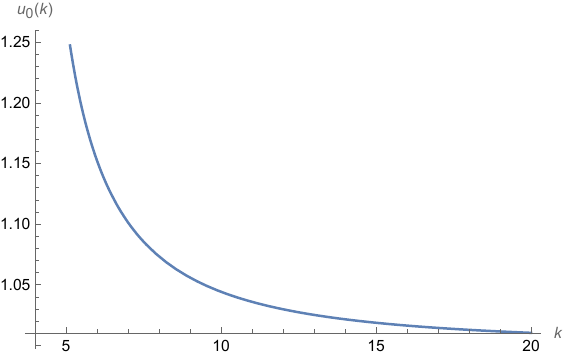}.
\end{equation}
In particular, we see that $u_0$ diverges when $k$ approaches to $4$. This in fact indicates a certain phase transition \cite{Bao:2021fjd}.
\end{example}

\subsection{Reflexive Polygons}\label{reflexive}
\begin{figure}[H]
    \centering
	\includegraphics[width=15cm]{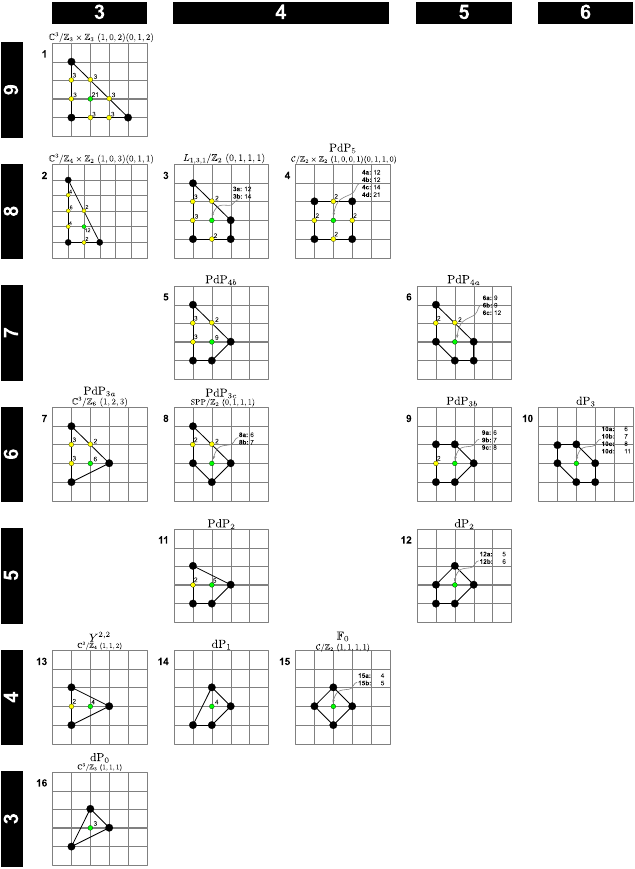}
	\caption{The 16 inequivalent reflexive polygons (up to SL$(2,\mathbb{Z})$). Figure taken from \cite{Hanany:2012hi} (with slight modifications). The reflexive polygons are arranged such that the dual pairs are mirror symmetric with respect to the middle line (fourth row), and the four polygons in the middle line are therefore self-dual. In each row, the polygons have the same number of boundary points/(normalized) area. In each column, the polygons have the same number of vertices.}\label{refpolygons}
\end{figure}

The next concept we will introduce is that of lattice polygons.
Given a Laurent polynomial $P=\sum c_{(m,n)}z^mw^n$, we can associate a point $(m,n)$ on the lattice $\mathbb{Z}^2$ to each monomial $z^mw^n$. In particular, we are interested in the convex hull of the set of these lattice points, which gives a (convex) lattice polygon called the  {\it Newton polygon} of $P$. In general, for an $n$-dimensional lattice polytope $\mathfrak{P}$, it is said to be \textbf{reflexive} if its dual polytope
\begin{equation}
    \mathfrak{P}^\circ:=\{\bm{v}\in\mathbb{Z}^n|\bm{u}\cdot\bm{v}\geq-1,\forall\bm{u}\in\mathfrak{P}\}
\end{equation}
is also a lattice polytope (in $\mathbb{Z}^n$). In two dimensions, a polygon is reflexive if and only if it has exactly one interior point. It is well-known that altogether there are 16 inequivalent reflexive polygons up to SL$(2,\mathbb{Z})$ transformations. These are listed in Figure \ref{refpolygons}.

\begin{example}
As the running example, we take $P=k-z-z^{-1}-w-w^{-1}$, which gives lattice points $(0,0)$, $(\pm1,0)$ and $(0,\pm1)$. Hence, it corresponds to a reflexive polygon with a single interior lattice point at $(0,0)$. This is called $\mathbb{F}_0$ (No.15) in Figure \ref{refpolygons}.
\end{example}

For each lattice polygon, we can always construct an associated toric variety \cite{fulton2016introduction} which is a complex surface.
For the reflexive ones, the variety will be Fano.
Furthermore, one can construct a
cone by viewing the points as $(m,n,1)$ in three dimensions and using $(0,0,0)$ as the apex. As a result, this defines a toric CY singularity of (complex) dimension 3 \cite{fulton2016introduction,cox2011toric}. Hence, the Newton polygon is also known as the toric diagram. As the endpoints $(m,n,1)$ of the cone are co-hyperplanar, the non-compact singularity is CY.

As shown in \cite{Hori:2000kt}, the Laurent/Newton polynomial $P(z,w)$ specifies the mirror geometry of the CY singularity by $P(z,w)=W=uv$ with $u,v\in\mathbb{C}$. Hence, it can be viewed as a double fibration over the $W$-plane. In particular, $P=0$ is known as the spectral curve. See \cite{Feng:2005gw} for more details.

\subsubsection{Tempered Polynomials}\label{tempered}
Given a Newton polygon, it is easy to construct the Newton polynomial, in reverse to what was discussed in the beginning of \S\ref{reflexive}. 
Nevertheless, we still have the freedom to choose the complex coefficients in $P(z,w)$. In number theory and related study of Mahler measure, the so-called tempered families are of particular interest \cite{villegas1999modular,boyd2002mahler}.

Given a Newton polygon $\mathfrak{P}$, we obtain the Newton polynomial $P(z,w)=\sum c_{(m,n)}z^mw^n$ with coefficients $c_{(m,n)}$ for each of the lattice points\footnote{Notice that the coefficients for the corner points should be non-zero.}. 
Now, consider a bounding edge $F$ of the polygon $\mathfrak{P}$. There might also be lattice points on it (the yellow points in Figure \ref{refpolygons}), in addition to the 2 endpoints (the black points in Figure \ref{refpolygons}) which are vertices of $\mathfrak{P}$.
Suppose there are $N$ lattice points on $F$, indexed from 0 to $N-1$, and we call the associated coefficients $c_{(m,n)}$ as $c_{F,l}$.
Then, we can create an auxiliary polynomial $P_F(t)\in\mathbb{C}[t]$ as 
\begin{equation}
    P_F(t)=\sum_{l=0}^{N-1}c_{F,l}t^l \ ,
\end{equation}
for each edge $F$.
  
Notice that this automatically requires that the boundary point $c_{F_1,N-1}$ to coincide with $c_{F_2,0}$ for any two adjacent edges $F_1$ and $F_2$. A Laurent polynomial is then said to be \textbf{tempered} if the set of roots of $\prod\limits_{F\in\mathfrak{P}}P_F(t)$ consists of roots of unity only. In other words, each $P_F$ in $\mathfrak{P}$ would only have roots on the unit circle.

Notice that being tempered only gives restrictions to coefficients for the boundary points. For the reflexive polygons considered in this paper, we always take the single interior point as the origin, corresponding to the constant term $k$ in the Newton polynomial as discussed before: 
$P(z,w) = k - p(z,w)$.

\begin{example}
For $\mathbb{F}_0$, $P=k-z-z^{-1}-w-w^{-1}$ is tempered. For instance, the lattice points $(1,0)$ and $(0,-1)$ corresponds to the monomials $-z$ and $-w^{-1}$ in $P$. 
The edge linking them is associated to the polynomial $-1-t$ which only has one root $t=-1$. In fact, every one of the 4 edges has the same polynomial $P_F=-1-t$. Thus, $P$ is tempered.
\end{example}

\begin{table}[h]
\centering
{\tiny
\begin{tabular}{|c|c|c|c|c|c|c|c|c|c|c|c|c|c|c|c|c|c|c|c|c|c|c|c|c|c|c|}
\hline
$t^0$ & 1 & 1 & 1 & 1 & 1 & 1 & 1 & 1 & 1 & 1 & 1 & 1 & 1 & 1 & 1 & 1 & 1 & 1 & 1 & 1 & 1 & 1 & 1 & 1 & 1 \\ \hline
$t^1$ & 0 & 0 & 0 & 0 & 0 & 0 & 0 & 0 & 0 & 1 & 1 & 1 & 1 & 1 & 1 & 1 & 1 & 2 & 2 & 2 & 2 & 2 & 3 & 3 & 4 \\ \hline
$t^2$ & $-2$ & $-1$ & $-1$ & 0 & 0 & 0 & 1 & 1 & 2 & $-1$ & 0 & 0 & 0 & 1 & 1 & 1 & 2 & 0 & 1 & 2 & 2 & 3 & 3 & 4 & 6 \\ \hline
$t^3$ & 0 & 0 & 0 & 0 & 0 & 1 & 0 & 0 & 0 & $-1$ & $-1$ & 0 & 1 & 0 & 1 & 1 & 1 & $-2$ & 0 & 1 & 2 & 2 & 1 & 3 & 4 \\ \hline
$t^4$ & 1 & 0 & 1 & $-1$ & 1 & 0 & 0 & 1 & 1 & 0 & $-1$ & 0 & 1 & 0 & 0 & 1 & 1 & $-1$ & 0 & 0 & 1 & 1 & 0 & 1 & 1 \\ \hline
\end{tabular}}
\caption{Each column gives a set of coefficients for $P_F(t)$ such that the solutions to $P_F(t)=0$ only have roots of unity.}\label{temperedlist}
\end{table}

For reflexive polygons, all the possible $P_F$'s that make $P$ tempered have been classified in \cite{villegas1999modular}. We reproduce it here in Table \ref{temperedlist}; there are 25 possibilities.
For convenience, given a tempered Newton polynomial, if it only has non-zero coefficients for vertices, we shall call such choice \emph{minimally tempered coefficients}. If all the boundary points have non-zero coefficients and the coefficients for every edge are binomial, that is, $P_F=(t+1)^N$ for all $F\in\mathfrak{P}$, then we shall call such choice \emph{maximally tempered coefficients}. When a polygon has no boundary lattice points other than vertices (i.e., each edge has exactly the 2 endpoints which are lattice points), the minimally and maximally tempered coefficients coincide and this is the only set of tempered coefficients.

\begin{example}
As we have seen, $\mathbb{F}_0$ has only one possible set of tempered coefficients. On the other hand, for $\mathbb{C}^3/(\mathbb{Z}_4\times\mathbb{Z}_2)$ $(1,0,3)(0,1,1)$ (No.2 in Figure \ref{refpolygons}), there are $2\times2\times4=16$ tempered choices. When all the three faces have $P_F(t)=-t-1$, $P(z,w)$ is minimally tempered. If $P_{F_{1,2}}=-t^2-2t-1=-(t+1)^2$ and $P_{F_3}=-t^4-4t^3-6t^2-4t-1=-(t+1)^4$, then $P(z,w)$ is maximally tempered. Notice that the minus sign is just a convention here as it would not change the spectral curve $P=0$. All the maximally and minimally tempered Newton polynomials are listed in Appendix \ref{Pzw}.
\end{example}

\begin{remark}
Returning to the Mahler measure,
we checked the Mahler flow for the $16$ reflexive polygons from $k=0$ to tropical $k$ numerically. The Mahler measures are all strictly increasing when $k$ increases, which supports the Mahler flow conjecture in \cite{Bao:2021fjd}.
\end{remark}

For future reference, let us list $u_0$ in Table \ref{refu0} for all the reflexive polygons with maximally tempered coefficients. As we can see, there is a grouping of the reflexive polygons in the sense that different polygons could have the same $u_0$ (and hence the Mahler measures). The discussions in \S\ref{modularity} would mainly follow this grouping.
\begingroup
\renewcommand{\arraystretch}{1.5}
\begin{longtable}{|c|c|} \hline
No.1 & $1+\frac{54}{k^2}+\frac{492}{k^3}+\frac{9882}{k^4}+\frac{158760}{k^5}+\frac{2879640}{k^6}+\frac{51982560}{k^7}$ \\
 & $+\frac{964347930}{k^8}+\frac{18091565520}{k^9}+\frac{343559141604}{k^{10}}+\dots$ \\ \hline
No.2, 3, 4 & $1+\frac{20}{k^2}+\frac{96}{k^3}+\frac{1188}{k^4}-\frac{10560}{k^5}+\frac{111440}{k^6}+\frac{1142400}{k^7}$ \\
 & $+\frac{12154660}{k^8}+\frac{130220160}{k^9}+\frac{1414339920}{k^{10}}+\dots$ \\ \hline
No.5, 6 & $1+\frac{10}{k^2}+\frac{30}{k^3}+\frac{270}{k^4}+\frac{1560}{k^5}+\frac{11350}{k^6}+\frac{77700}{k^7}$ \\
 & $+\frac{560350}{k^8}+\frac{4040400}{k^9}+\frac{29623860}{k^{10}}+\dots$ \\ \hline
No.7, 8, 9, 10 & $1+\frac{6}{k^2}+\frac{12}{k^3}+\frac{90}{k^4}+\frac{360}{k^5}+\frac{2040}{k^6}+\frac{10080}{k^7}$ \\
 & $+\frac{54810}{k^8}+\frac{290640}{k^9}+\frac{1588356}{k^{10}}+\dots$ \\ \hline
No.11, 12 & $1+\frac{4}{k^2}+\frac{6}{k^3}+\frac{36}{k^4}+\frac{120}{k^5}+\frac{490}{k^6}+\frac{2100}{k^7}$ \\
 & $+\frac{8260}{k^8}+\frac{36960}{k^9}+\frac{151704}{k^{10}}+\dots$ \\ \hline
No.13, 15 & $1+\frac{4}{k^2}+\frac{36}{k^4}+\frac{400}{k^6}+\frac{4900}{k^8}+\frac{63504}{k^{10}}+\dots$ 
  $={}_2F_1\left(\frac{1}{2},\frac{1}{2};1;16k^{-2}\right)$ \\ \hline
No.14 & $1+\frac{2}{k^2}+\frac{6}{k^3}+\frac{6}{k^4}+\frac{60}{k^5}+\frac{110}{k^6}+\frac{420}{k^7}$ \\
 & $+\frac{1750}{k^8}+\frac{4200}{k^9}+\frac{19152}{k^{10}}+\dots$ \\ \hline
No.16 & $1+\frac{6}{k^3}+\frac{90}{k^6}+\frac{1680}{k^9}+\dots$
  $={}_2F_1\left(\frac{1}{3},\frac{2}{3};1;27k^{-3}\right)$ \\ \hline
\caption{The periods $u_0$ to order 10 for reflexive polygons with maximally tempered coefficients.}\label{refu0}
\end{longtable}
\endgroup

\subsubsection{Elliptic Curves}\label{elliptic}
Since the reflexive polygons give elliptic curves, we here review some of the requisites from the geometry and number theory of elliptic  curves.
In general, any elliptic curve $E$ can be transformed into Weierstrass normal form
\begin{equation}
    y^2=x^3+fx+g.
\end{equation}
The curve is non-singular if and only if $\Delta\neq0$, where
\begin{equation}
    \Delta=-16(4f^3+27g^2)
\end{equation}
is known as the discriminant. Then the $j$-invariant is given by
\begin{equation}
    j=-\frac{2\times(24f)^3}{\Delta}.
\end{equation}
This is a crucial concept since isomorphic (isogenous) elliptic curves have the same $j$-invariant. Notice that however $j$-invariant is only able to distinguish elliptic curves over algebraically closed fields.

Topologically, an elliptic curve $E$ is the torus $\mathbb{T}^2$. Hence, it is endowed with a complex structure specified by the two periods which are integrals along the two cycles $A$ and $B$ of the torus: $\int_{A,B}\frac{\text{d}x}{y}$. This complex structure should coincide with the $\tau$ computed from $u_{0,1}$ in \eqref{def:tau} up to SL$(2,\mathbb{Z})$. As a function of $\tau$, $j(\tau):\mathfrak{h}\rightarrow\mathbb{P}^1$ is a modular function, i.e., invariant under SL$(2,\mathbb{Z})$ transformations.
It is in fact the only modular function in that any meromorphic function which is SL$(2,\mathbb{Z})$-invariant is a rational function in $j(\tau)$.

Now, because our Newton polynomial \eqref{kNP} always has a parameter $k$, any reflexive polygon defines for us a family of elliptic curves.
Geometrically, when $k \in \mathbb{C} \sqcup \infty$, this defines an elliptic fibration over $\mathbb{P}^1$, giving us a complex surface which is called an elliptic surface \cite{shioda1972elliptic,He:2012jn}.
In this case, all the crucial quantities, such as $\Delta$ and $j$, depend on $k$. In particular, $j(k)$ can be seen as a map from $\mathbb{P}^1$ with coordinate $k$ to $\mathbb{P}^1$. We will make use of this map shortly.

\subsection{Modular Mahler Measure}\label{modularmahler}
In general, the spectral curve $P(z,w) = 0$ defines a Riemann surface as an algebraic curve $\Sigma$. Since each reflexive polygon has a single interior point, $\Sigma$ is genus one. For all but finitely many $k$'s, the curve would be a smooth elliptic curve.
For convenience, let us define $\lambda:=k^{-1}$, then we have (where we explicitly write out the dependence of the elliptic curve on the parameter $\lambda$)
\begin{equation}
    \Sigma_{\lambda} \ : \
    1-\lambda \ p(z,w) = 0 \ .
\end{equation}
As pointed out in \cite{villegas1999modular}, $u_0$ is a period of a holomorphic 1-form on $\Sigma_\lambda$. Hence, it satisfies the Picard-Fuchs equation
\begin{equation}
    A(\lambda)\frac{\text{d}^2u_0}{\text{d}\lambda^2}+B(\lambda)\frac{\text{d}u_0}{\text{d}\lambda}+C(\lambda)u_0=0,\label{PF}
\end{equation}
where $A(\lambda),B(\lambda),C(\lambda)$ are polynomials in $\lambda$. 
We may then use the Picard-Fuchs equation to find the dual period $u_1$ of the form
\begin{equation}
    u_1(\lambda)=u_0(\lambda)\log(\lambda)+v(\lambda),
\end{equation}
where $v$ is a holomorphic function with $v(0)=0$. This defines 
\begin{equation}\label{def:tau}
    \tau=\frac{1}{2\pi i}\frac{u_1}{u_0},\quad q=\text{e}^{2\pi i\tau}=\lambda+\dots.
\end{equation}
As usual, $\tau$ gives the complex structure of the elliptic curve $P=0$ as a torus. The monodromy around $\lambda=0$ (i.e., at $k$ infinity) acts as $\tau\rightarrow\tau+1$. This fixes $q$ and we may locally invert it to get
\begin{equation}
    \lambda(\tau)=q+\dots,\quad u_0(\tau)=1+\dots.
\end{equation}
Using the nome $q$, we can also express the Mahler flow equation \eqref{mahlerfloweqn} as
\begin{equation}
    q\frac{\text{d}m}{\text{d}q}=\frac{\text{d}m}{\text{d}\lambda}\frac{q\text{d}\lambda}{\text{d}q}=\frac{u_0}{\lambda}\frac{q\text{d}\lambda}{\text{d}q}=:e(\tau).\label{etau}
\end{equation}

In fact, $\lambda,u_0,e$ are modular forms (with singularities) of weights 0, 1, 3 respectively under the monodromy of Picard-Fuchs equation, namely a congruence subgroup of SL$(2,\mathbb{Z})$ acting on $\tau$  \cite{villegas1999modular}. We may therefore call \eqref{etau} the \textbf{modular Mahler flow equation}.

Write the Fourier series of $e(\tau)$ as $e(\tau)=1+\sum\limits_{n=1}^\infty e_nq^n$. Then from \eqref{etau}, we have
\begin{theorem}[Rodriguez-Villegas \cite{villegas1999modular}]
Locally around $\tau=i\infty$ (i.e., $\lambda=0$), we have
\begin{equation}
    m(P)=-2\pi i\tau-\sum_{n=1}^\infty\frac{e_n}{n}q^n.
\end{equation}
\end{theorem}
Because of the modularity of $e(\tau)$, the Mahler measure for elliptic curves is referred to as {\it modular Mahler measure} though $m(P)$ itself is not modular.

\begin{example}
For $P(z,w)=\lambda^{-1}-z-z^{-1}-w-w^{-1}$, $m(P)$ and $u_0$ are given in \eqref{mu0ex}. Since $u_0$ is hypergeometric, it is easy to see that the Picard-Fuchs equation is
\begin{equation}
    \mu(16\mu-1)\frac{\textup{d}^2u}{\textup{d}\mu^2}+(32\mu-1)\frac{\textup{d}u}{\textup{d}\mu}+4u=0,
\end{equation}
where we have used $\mu:=\lambda^2$ for convenience. This leads to \cite{villegas1999modular}
\begin{equation}
    u_1=u_0\log(\mu)+8\mu+84\mu^2+\frac{2960}{3}\mu^3+\dots,
\end{equation}
and
\begin{equation}
    u_0=1+4\sum_{n=1}^\infty\sum_{d|n}\chi_{-4}(d)q^n,\quad e=1-4\sum_{n=1}^\infty\sum_{d|n}\chi_{-4}(d)d^2\,q^n,\quad\mu=\frac{1}{c^2}\left(\sum_{\substack{n=1\\n\textup{ odd}}}^\infty\sum_{d|n}d\,q^n\right),
\end{equation}
where $\chi_{-4}$ is the Dirichlet character/Kronecker symbol satisfying $\chi_{-4}(n)=1,0$ when $n\equiv0,1~(\textup{mod }2)$. Then, we have
\begin{equation}
    m(P)=\frac{16\,\textup{Im}\tau}{\pi^2}\sum_{\substack{n_1,n_2\in\mathbb{Z}\\(n_1,n_2)\neq(0,0)}}\frac{\chi_{-4}(n_1)}{(n_1+4n_2\tau)^2(n_1+4n_2\Bar{\tau})},
\end{equation}
which under modular transformations, we have
\begin{equation}
    \begin{split}
        &\tau\rightarrow-\frac{1}{\tau}:\quad m=\frac{16\,\textup{Im}\Bar{\tau}}{\pi^2}\sum_{\substack{n_1,n_2\in\mathbb{Z}\\(n_1,n_2)\neq(0,0)}}\frac{\chi_{-4}(n_1)\tau}{(4n_2-n_1\tau)^2(4n_2-n_1\Bar{\tau})},\\
        &\tau\rightarrow\tau+1:\quad  m=\frac{16\,\textup{Im}\tau}{\pi^2}\sum_{\substack{n_1,n_2\in\mathbb{Z}\\(n_1,n_2)\neq(0,0)}}\frac{\chi_{-4}(n_1)}{(n_1+4n_2\tau)^2(n_1+4n_2\Bar{\tau})}.
    \end{split}
\end{equation}
Hence, $m$ is invariant under monodromy ($\tau\rightarrow\tau+1$) at $k\rightarrow\infty$ while we have $m\rightarrow0$ as $\tau\rightarrow0$. We expect this to be true in general for reflexive polygons.
\end{example}

\subsection{Esquisse de Dessins, or Sketches of Children's Drawings}\label{sketch}
The discussions on elliptic curves above are initmately related to the profound theorem by Belyi \cite{belyui1980galois}:
\begin{theorem}
Let $\mathcal{X}$ be a compact, connected Riemann surface. Then $\mathcal{X}$ is a non-singular, irreducible projective variety of complex dimension $1$ and can be defined by polynomial equations. The defining polynomial has algebraic coefficients if and only if there exists a rational map $\beta:\mathcal{X}\rightarrow\mathbb{P}^1$ which is ramified at exactly three points, that is, has three critical values.
\end{theorem}

We will be primarily concerned with the case of $\mathcal{X} = \mathbb{P}^1$, so that the Belyi map is a rational function $p(x)/q(x) : \mathbb{P}^1 \rightarrow \mathbb{P}^1$. 
Now, on the target $\mathbb{P}^1$, any three points can be taken to be 0, 1 and $\infty$ (that is, $[0:1]$, $[1:1]$ and $[1:0]$ in homogenous coordinates) by linear-fractional M\"obius transformations, so that the three ramified points can be thus chosen.
Following Grothendieck \cite{Grothendieck1984sketch}, a bipartite graph called \textbf{dessin d'enfant} (or child's drawing) can be associated to $\beta$ by
\begin{equation}
    \begin{split}
        &B=\beta^{-1}(0)=\{x\in\mathbb{P}^1~|~p(x)=0\},\quad W=\beta^{-1}(1)=\{x\in\mathbb{P}^1~|~p(x)=q(x)\},\\
        &E=\beta^{-1}([0,1])=\{x\in\mathbb{P}^1~|~p(x)=tq(x),\text{ for some $t\in[0,1]$}\},\\
        &F=\beta^{-1}(\infty)=\{x\in\mathbb{P}^1~|~q(x)=0\},
    \end{split}
\end{equation}
where $B$, $W$, $E$ and $F$ denote the black, white vertices, edges and faces respectively. As $\beta$ is $\mathbb{P}^1\rightarrow\mathbb{P}^1$, the graph is embedded on a sphere. Moreover,
\begin{proposition}
Let $\beta:\mathbb{P}^1\rightarrow\mathbb{P}^1$ be a Belyi map. Then the associated bipartite graph $(V=B\sqcup W,E,F)$ is loopless, connected and planar. It has $|V|=|\beta^{-1}(\{0,1\})|$ vertices, $|E|=\deg(\beta)$ edges and $|F|=|\beta^{-1}(\infty)|$ faces, satisfying $|V|-|E|+|F|=2$.
\end{proposition}
As we will plot the dessin on a plane via stereographic projection, all the bounded faces on the plane are called internal faces while the face containing $x\rightarrow\infty$ is known as the external face. As $\beta$ is a multi-covering of the target $\mathbb{P}^1$, we can consider the monodromy around each vertex in the dessin. Essentially, each monodromy acting on a vertex permutes the edges connected to that vertex. We shall denote the set of such permutations around black (white) vertices as $\sigma_0$ ($\sigma_1$). Then $\sigma_0$ and $\sigma_1$ generate a free group known as the monodromy/cartographic group $G$ of the dessin. In particular, the monodromies $\sigma_{\infty}$ around faces can be obtained by $\sigma_\infty\circ\sigma_1\circ\sigma_0=1$. As the dessin has $|E|$ edges, $G$ is a subgroup of the symmetric group $\mathfrak{S}_N$ where $N=|E|$.

In our context, recall that all our elliptic curves $E$ are parametrized by $k$ so that the Klein invariant $j(k)$ is a function of the parameter $k$ and is thus a map from $\mathbb{P}^1$ (instead of $E$) to $\mathbb{P}^1$.
We will show in \S\ref{dessinsMahler}
that $j(k)$
is actually Belyi for maximally tempered coefficients in the Newton polynomials:
\begin{equation}
    \beta=\frac{j}{1728} \ .
\end{equation}

\paragraph{Congruence subgroups and coset graphs} A coset graph is a graph associated with a group $K$ generated by elements $\{x_i\}$ and a subgroup $H$. Then each vertex (drawn in black so as to reconstruct the dessin) in the coset graph represents a right coset $Hg$ for $g\in K$. An edge is of form $(Hg,Hgx_i)$ which connects the coset $Hg$ and $H(gx_i)$.

As we will see shortly, the dessins associated to reflexive polygons (with maximally tempered coefficients) are \emph{clean}, namely that the white vertices all have valency $2$. Therefore, the dessins can be viewed as coset graphs by removing the white vertices. Conversely, we can insert a white vertex on each edge to get the dessin from the coset graph.

In particular, the dessins we will consider in \S\ref{modularity} are associated with the modular group (P)SL$(2,\mathbb{Z})$ and the congruence subgroups. Hence, the generators can be taken as the usual $S$ and $T$, viz,
\begin{equation}
    S=\begin{pmatrix}
0 & -1 \\
1 & 0
\end{pmatrix},\quad
 T=\begin{pmatrix}
1 & 1 \\
0 & 1
\end{pmatrix}\quad
\text{s.t. PSL}(2,\mathbb{Z})=\langle S,T|S^2=(ST)^3=1\rangle.
\end{equation}
The \textbf{congruence groups} of level $n$ are defined as
\begin{equation}
    \begin{split}
        &\Gamma(n):=\left\{M\in\text{(P)SL}(2,\mathbb{Z})\Bigg|M\equiv\begin{pmatrix}
1 & 0 \\
0 & 1
\end{pmatrix}~(\text{mod }n)\right\},\\
&\Gamma_1(n):=\left\{M\in\text{(P)SL}(2,\mathbb{Z})\Bigg|M\equiv\begin{pmatrix}
1 & b \\
0 & 1
\end{pmatrix}~(\text{mod }n)\right\},\\
&\Gamma_0(n):=\left\{M\in\text{(P)SL}(2,\mathbb{Z})\Bigg|M\equiv\begin{pmatrix}
a & b \\
0 & d
\end{pmatrix}~(\text{mod }n)\right\}.
    \end{split}
\end{equation}
In particular, we have $\Gamma(n)\leq\Gamma_1(n)\leq\Gamma_0(n)$ and $\Gamma_0(n_2)\leq\Gamma_0(n_1)$ if $n_1|n_2$.
The fact that every congruence subgroup of (P)SL$(2,\mathbb{Z})$ has a coset graph  (called Schreier-Cayley graph) which is a clean trivalent dessin was discussed in \cite{He:2012kw,He:2012jn,Tatitscheff:2018aht}.

Given a congruence subgroup $\Gamma$, the quotient space $\mathfrak{h}/\Gamma$ (where $\mathfrak{h}$ is the upper half plane) can be compactified by adding a few isolated points (aka cusps of $\Gamma$). Such compactified curve $X(\Gamma)$ is called the \textbf{modular curve}. The genus of $\Gamma$ is then defined to be the genus of $X(\Gamma)$. When $X(\Gamma)$ is of genus 0, the field of meromorphic functions on $X(\Gamma)$ is generated by a single element known as a \textbf{Hauptmodul} of $\Gamma$.

\subsection{Dimers and Quivers}\label{dimers}
Having introduced the requisite mathematical concepts, let us now proceed to some physics.
With so-called forward and inverse algorithms, one can construct a Newton polygon, which encodes a toric CY 3-fold, from a brane tiling, which encodes a quiver gauge theory in (3+1)-dimensions with $\mathcal{N}=1$ supersymmetry, and vice versa. 
From a tiling, we may then get a toric quiver as its dual graph which encodes certain supersymmetric gauge theories. Here, we will not explain the details and the readers are referred to \cite{Feng:2000mi,Hanany:2005ve,Franco:2005rj,Franco:2005sm,Feng:2005gw}. Instead, we will only introduce the necessary concepts for this paper.

A {\bf perfect matching} of a bipartite graph is a collection of edges such that each vertex is incident to exactly one edge.
The bipartite nature then means that each edge connects one white with one black node.
The \textbf{dimer model} is then the study of (random) perfect matchings \cite{Kenyon:2003uj,kenyon2003introduction}. Physically, the dimer models are also called \textbf{brane tilings} as they have a nice interpretation in terms of brane systems. In particular, the dimers discussed here are all $\mathbb{Z}^2$-periodic and embedded in $\mathbb{T}^2$.

Given a perfect matching $M$, there is a unit flow $\omega$ along each edge in $M$ from a white to a black node. Consider a reference perfect matching $M_0$ with flow $\omega_0$ and a path from face $f_0$ to $f_1$ in the graph. Then for any $M$ with flow $\omega$, the total flux of $\omega-\omega_0$ across the path defines a path-independent height function of $M$. The difference of height functions of any two perfect matchings is independent of the choice of $M_0$. The height change is then $(h_x,h_y)$ if the horizontal and vertical height changes of $M$ (for one period) are $h_x$ and $h_y$ respectively.

Moreover, we can assign some (real) energy function to each edge $e$. Then given a perfect matching $M$, its energy is $\mathcal{E}(M):=\sum\limits_{e\in M}\mathcal{E}(e)$. For any edge $e$ in the graph, its \textbf{edge weight} is defined to be $\textup{e}^{-\mathcal{E}(e)}$.

With all these quantities, we can then obtain the associated Newton polynomial as \cite{Kenyon:2003uj}
\begin{equation}
    P(z,w)=\sum_M\text{e}^{-\mathcal{E}(M)}z^{h_x}w^{h_y}(-1)^{h_xh_y+h_x+h_y}.\label{dimer2polynomial}
\end{equation}
Moreover, from this expression, we can see that each perfect matching $M$ is associated to a lattice point $(h_x,h_y)$ in the Newton polygon. Hence, we shall call a perfect matching \textbf{external} (\textbf{internal}) if it corresponds to an external (internal) point in the toric diagram. Physically, the perfect matchings are in one-to-one correspondence with gauged linear sigma model fields. The perfect matchings associated to each point are explicitly labelled in Figure \ref{refpolygons} for reflexive polygons. Since the polygon/tiling correspondence is one-to-many, the different numbers assigned to the interior point of a polygon are from different tilings/toric phases. Notice that each vertex corresponds to exactly one perfect matching.

\paragraph{Canonical edge weights} There is a special choice of edge weights which comes from the R-charges of the fields in the quiver theory. Since the quiver is the dual diagram, each arrow\footnote{We can choose the direction of an arrow such that the white (or black) node is on its left as long as the convention is consistent.} corresponds to an edge in the tiling. As each arrow is physically a multiplet, we can relate its R-charge $R$ to the corresponding edge in the tiling as follows. As an edge always belongs to two faces, we can connect the centre points of the faces and the two endpoints of the edge to form a rhombus. Then the rhombus angle $\theta$ satisfies $2\theta=\pi R$, where $2\theta$ is the angle of the rhombus at the vertex that has in common with the edge. 

In particular, the edge would then have length $2\cos(\theta)$. This is often the situation for isoradial dimers whose faces can all be inscribed in a circle of the same radius. Nevertheless, as discussed in \cite{Bao:2021fjd}, this R-charge/rhombus angle setting could also be extended to non-isoradial dimers if one thinks of zero or negative edge lengths. Then for convenience, we say that the \textbf{canonical weight} of an edge is $2\sin(\pi R/2)$. With such canonical weights, one can find many interesting features of quiver gauge theories regarding the Newton polynomials and Mahler measures. See \cite{Bao:2021fjd} for more details.
The point is that with the canonical edge weights gives a natural way of fixing the coefficients in a Newton polynomial.

\begin{example}
Let us consider the tiling in Figure \ref{dimerexample}(a). 
\begin{figure}[h]
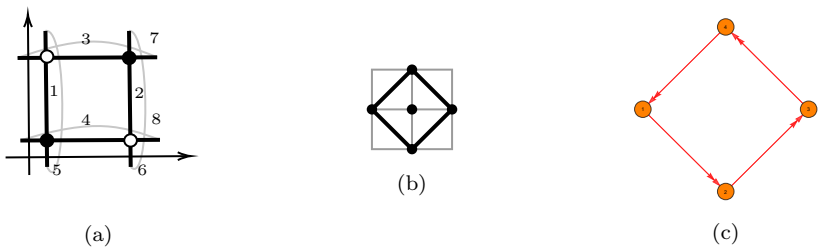

    \centering
    \include{tilingF0example}
    \caption{(a) The dimer model. (b) The toric diagram. (c) The quiver diagram.}\label{dimerexample}
\end{figure}
The canonical weight for an edge is always $\sqrt{2}$. Using \eqref{dimer2polynomial}, one can find that the curve is given by
\begin{equation}
    -2z-2z^{-1}-2w-2w^{-1}+(2+2+2+2)=0,
\end{equation}
or, equivalently,
\begin{equation}
    -z-z^{-1}-w-w^{-1}+4=0.
\end{equation}
We see that the corresponding Newton polygon is our running reflexive example, shown in \ref{dimerexample}(b).
The quiver in Figure \ref{dimerexample}(c) is the dual graph of the dimer.
This is the well-known phase I (corresponding to No.15 (a) in Figure \ref{refpolygons}) of the D-brane worldvolume quiver gauge theory for the CY 3-fold which is the cone over the Fano toric surface $\mathbb{F}_0 = \mathbb{P}^1 \times \mathbb{P}^1$.
In this example, we find that Newton polynomial obtained from canonical weights coincide with the one with (maximally) tempered coefficients (at $k=4$).

In general, coefficients from canonical weights are not necessarily the same as maximally tempered coefficients.
\end{example}

\paragraph{Specular duality} As studied in \cite{Hanany:2012vc}, the quiver gauge theories enjoy \textbf{specular duality} under which the master spaces are preserved. In short, the master space $\mathcal{F}^{\flat}$ \cite{Forcella:2008eh,Forcella:2008bb} is the space of solutions to the F-term relations (more strictly, the maximal spectrum of the coordinate ring quotiented by the ideal of F-terms). Its largest irreducible component is known as the coherent component ${}^\text{Irr}\mathcal{F}^{\flat}$, which can be written as a symplectic quotient in terms of perfect matchings $M_i$:
\begin{equation}
    {}^\text{Irr}\mathcal{F}^{\flat}=\mathbb{C}[M_i]//Q_F,
\end{equation}
where $Q_F$ is the F-term charge matrix \cite{Feng:2000mi}.

\begin{remark}
It was shown in \cite{Bao:2021fjd} that $u_0(k)$ is the generating function of the master space in terms of F-term charge matrix though the F-term relations in higher orders are just redundant. Following \cite{Bao:2021fjd}, $k\rightarrow\infty$ is known as a \textbf{tropical limit} when $P(z,w)=k-p(z,w)$. As a result, this is equivalent to $q\rightarrow0$ and hence $\tau\rightarrow i\infty$. Physically, $\tau$ can be interpreted as the complexified gauge coupling in type IIB string theory, that is, $\tau=\frac{\theta}{2\pi}+\frac{i}{g_{\textup{IIB}}}$. Therefore, this gives the weak coupling $g_\textup{IIB}\rightarrow0$. For the modular forms (with singularties) introduced before, we have $\lambda\rightarrow0$, $u_0\rightarrow1$ and $e\rightarrow1$. In particular, $u_0\rightarrow1$ indicates a free theory in the tropical limit as the master space become trivial. This is consistent with the weakly coupled gauge theory with $\tau\rightarrow i\infty$.

Moreover, we have $m$ tends to $\infty$ for tropical $k$. On the other hand, $m$ would go to $0$ in the strong coupling regime (as $\tau\rightarrow0$).
\end{remark}

For reflexive polygons, the specular duals are \cite{Hanany:2012vc}
\begin{eqnarray}
&
1 \leftrightarrow 1~,~
&
\nn\\
&
2 \leftrightarrow 4d
~,~
3a \leftrightarrow 4c
~,~
3b \leftrightarrow 3b
~,~
4a \leftrightarrow 4a
~,~
4b \leftrightarrow 4b~,~
&
\nn\\
&
{\color{red} 5} \leftrightarrow {\color{red}6}c
~,~
6a \leftrightarrow 6a
~,~
6b \leftrightarrow 6b~,~
&
\nn\\
&
7 \leftrightarrow 10d
~,~
8a \leftrightarrow 10c
~,~
8b \leftrightarrow {\color{red}9}c
~,~
{\color{red} 9}a \leftrightarrow 10b
~,~
9b \leftrightarrow 9b
~,~
10a \leftrightarrow 10a~,~
&
\\
&
{\color{red} 11} \leftrightarrow 12b
~,~
12a \leftrightarrow 12a~,~
&
\nn\\
&
13 \leftrightarrow 15b
~,~
14 \leftrightarrow 14
~,~
15a \leftrightarrow 15a~,~
&
\nn\\
&
16 \leftrightarrow 16~,
&
\nn\label{specpairs}
\end{eqnarray}
where the letters following the number label different toric phases as in Figure \ref{refpolygons}. As we can see, their internal and external perfect matchings get exchanged under specular duality.

\begin{remark}
The red numbers in the above list contain polygons that are called exceptional cases due to the following two reasons \cite{Bao:2021fjd}. 
\begin{itemize}
    \item With canonical weights, most of the specular duals can have the same Mahler measure (with the constant term taken to be $k$) up to an additive constant, that is, $m(P_2)=m(P_1)+\kappa$. Hence, by an overall scaling of factor $\textup{e}^{-\kappa}$, we have $k-p_1\rightarrow k-\textup{e}^{-\kappa}p_1$ (notice that this does not change the spectral curve $P_1=0$). Thus, $m(P_2)=m(P_1)$. As a result, for example, polygons No.2 and No.3 have the same Mahler measure as they are connected by different toric phases of No.4 even though they are not specular duals. However, it turns out that No.5, 6, 9 and 11 do not satisfy this property.

\item Moreover, with canonical weights (with constant term $k$), most Newton polynomials can have equal coefficients for vertices under rescaling of $z$ and/or $w$. However, this is not possible for No.5, 6, 9 and 11.
\end{itemize}
It is worth noting that the exceptions of the above two properties coincide (though the reason why they coincide is still unclear). It is still not known why No.5, 6, 9, 11 are exceptional. In \S\ref{dessinsMahler}, we will see that they are further exceptional regarding a third property.\label{exceptions}
\end{remark}

\begin{example}
The toric phase (b) for No.15 is specular dual to the single toric phase for No.13. Indeed, No.15(b) has $4$ external and $5$ internal perfect matchings while No.13 has $5$ external and $4$ internal perfect matchings. A more detailed analysis on the two theories under specular duality can be found in \cite{Hanany:2012vc}.
\end{example}

\section{Modularity and Gauge Theories}\label{modularity}
Having introduced all the background, we are now ready to discuss how modular Mahler measures connected the various different areas in mathematics and physics. From \S\ref{background}, the readers may have already noticed that
\begin{proposition}
The maximally tempered coefficients in the Newton polynomials are equal to the numbers of perfect matchings associated to the exterior lattice points of the toric diagrams.
\end{proposition}
Hence, we will mainly focus on maximally tempered coefficients in the following discussions, and we will see various properties implying potential physical relevance. As listed in \cite{Bao:2020kji}, all the non-reflexive polygons with two interior points also have maximally tempered coefficients equal to the numbers of perfect matchings associated to the boundary points (it would also be interesting to see what happens for higher dimensional reflexive polytopes \cite{He:2017gam}). Furthermore, the consistent brane tilings for all polygons presented in \cite{Davey:2009bp} also have maximally tempered coefficients while the remaining inconsistent tilings do not\footnote{See \cite{Hanany:2015tgh} for a general discussion on consistency of brane tilings.}. Therefore, it is natural to conjecture that
\begin{conjecture}
A brane tiling is consistent if and only if the corresponding toric diagram (either reflexive or non-reflexive) has maximally tempered coefficients for its boundary points. Moreover, the maximally tempered coefficients are equal to the numbers of perfect matchings associated to the boundary points.
\end{conjecture}
It is curious that maximal tempered coefficients appear in two completely different contexts, one from perfect matching in physics and another from considering Mahler measure in mathematics.

\subsection{Dessins and Mahler Measure}\label{dessinsMahler}
As mentioned throughout, we will focus on the 16 reflexive polygons with maximally tempered coefficients. The Newton polynomials are listed in Table \ref{Pzwmaxmin}. Recall that the spectral curve $P(z,w)=0$ for each reflexive polygon is an elliptic curve (except for finitely many $k$ values). We can transform the spectral curves into Weierstrass normal form $y^2=x^3+f(k) x+g(k)$ (recall that all our elliptic curves depend on the parameter $k$). This is computationally rather involved (Nagell's algorithm) but can luckily be done with \texttt{SAGE} \cite{sage}.

In Table \ref{refelliptic}, we list the Weierstrass form of all 16 reflexive polygons with maximally tempered coefficients, where the coefficients $f(k)$ and $g(k)$ all assume the form
\begin{equation}
    f=-\frac{1}{48}k^4+a(k),\quad g=\frac{1}{864}k^6+b(k).
\end{equation}

\begingroup
\renewcommand{\arraystretch}{1.5}
\begin{longtable}{c||c|c} \hline
Polygon(s) & No.1 & No.2, 3, 4 \\ \hline
$a(k)$ & $\frac{9}{2}k^2+36k+81$ & $\frac{5}{3}k^2+8k+\frac{32}{3}$ \\ \hline
$b(k)$ & $-\frac{3}{8}k^4-4k^3-\frac{27}{2}k^2+54$ & $-\frac{5}{36}k^4-\frac{2}{3}k^3+\frac{8}{9}k^2+\frac{32}{3}k+\frac{448}{27}$ \\ \hline
$\Delta$ & $(k-21)(k+6)^8$ & $(k-12)(k+4)^7$ \\ \hline
Singular $k$ & $-6$, $21$ & $-4$, $12$ \\ \hline\hline
Polygon(s) & No.5, 6 & No.7, 8, 9, 10 \\ \hline
$a(k)$ & $\frac{5}{6}k^2+\frac{5}{2}k+\frac{5}{3}$ & $\frac{1}{2}k^2+k$  \\ \hline
$b(k)$ & $-\frac{5}{72}k^4-\frac{5}{24}k^3+\frac{5}{9}k^2+\frac{19}{6}k+\frac{395}{108}$ & $-\frac{1}{24}k^4-\frac{1}{12}k^3+\frac{1}{4}k^2+k+1$ \\ \hline
$\Delta$ & $(k+3)^5(k^2-5k-25)$ & $(k-6)(k+2)^3(k+3)^2$ \\ \hline
Singular $k$ & $-3$, $\frac{5}{2}(1\pm\sqrt{5})$ & $-3$, $-2$, $6$ \\ \hline\hline
Polygon(s) & No.11, 12 & No.13, 15 \\ \hline
$a(k)$ & $\frac{1}{3}k^2+\frac{1}{2}k-\frac{1}{3}$ & $\frac{1}{3}k^2-\frac{1}{3}$ \\ \hline
$b(k)$ & $-\frac{1}{36}k^4-\frac{1}{24}k^3+\frac{5}{36}k^2+\frac{1}{3}k+\frac{35}{108}$ & $-\frac{1}{36}k^4+\frac{5}{36}k^2+\frac{2}{27}$ \\ \hline
$\Delta$ & $(k+1)^2(k^3+k^2-18k-43)$ & $k^2(k^2-16)$ \\ \hline
Singular $k$ & $-1$, $\kappa_{1,2,3}$ & $0$, $\pm4$ \\ \hline\hline
Polygon(s) & No.14 & No.16 \\ \hline
$a(k)$ & $\frac{1}{6}k^2+\frac{1}{2}k-\frac{1}{3}$ & $\frac{1}{2}k$ \\ \hline
$b(k)$ & $-\frac{1}{72}k^4-\frac{1}{24}k^3+\frac{1}{18}k^2+\frac{1}{6}k+\frac{19}{108}$ & $-\frac{1}{24}k^3+\frac{1}{4}$ \\ \hline
$\Delta$ & $k^4+k^3-8k^2-36k-11$ & $k^3-27$ \\ \hline
Singular $k$ & $\kappa_{5,6,7,8}$ & $3$, $\frac{3}{2}(-1\pm\text{i}\sqrt{3})$ \\ \hline
\caption{The data of the elliptic curves for reflexive polygons with maximally tempered coefficients. We also list the values of $k$ when the spectral curve becomes singular for each case. Here, $\kappa_{1,2,3}$ are the three roots to $k^3+k^2-18k-43=0$ ($\kappa_1\approx4.73$, $\kappa_{2,3}\approx-2.86\pm0.94\text{i}$) while $\kappa_{5,6,7,8}$ are the four roots to $k^4+k^3-8k^2-36k-11=0$ ($\kappa_5\approx-0.33$, $\kappa_6\approx3.80$, $\kappa_{7,8}\approx-2.23\pm1.94\text{i}$).}\label{refelliptic}
\end{longtable}
\endgroup

We find that specular duals have exactly the same elliptic curve. Notice that this property only holds for maximally tempered coefficients\footnote{In Appendix \ref{minimally}, for example, we list the elliptic curves for the same polygons but with minimally tempered coefficients, and specular duals do not give the same elliptic curves anymore.}. Recall that the maximally tempered coefficients indicate the number of perfect matchings for each lattice point and that specular duality exchange internal and external perfect matchings. Again, we see that maximally tempered coefficients are of particular physical interest.

We also tabulate all the values of $k$ that make each spectral curve $P=0$ singular in Table \ref{refelliptic}. They can be obtained by checking whether the discriminant of the curve vanishes. It is worth mentioning that in many cases, there exists a singular $k$ such that $|k|$ is equal to the minimal number of internal perfect matchings for the polygon. For instance, No.4 has four toric phases, the numbers of internal perfect matchings are 12, 12, 14 and 21 respectively. Indeed, there is a singular $k=|k|=12$. However, five of the reflexive polygons do not obey this observation: No.~5, 6, 11, 14 and 12. We find that the first four polygons coincide with the exceptional cases in Remark \ref{exceptions} while No.12 is the specular dual of (the exceptional) No.11.

\begin{figure}[H]
    \centering
    \begin{subfigure}{6cm}
		\centering
		\includegraphics[width=5cm]{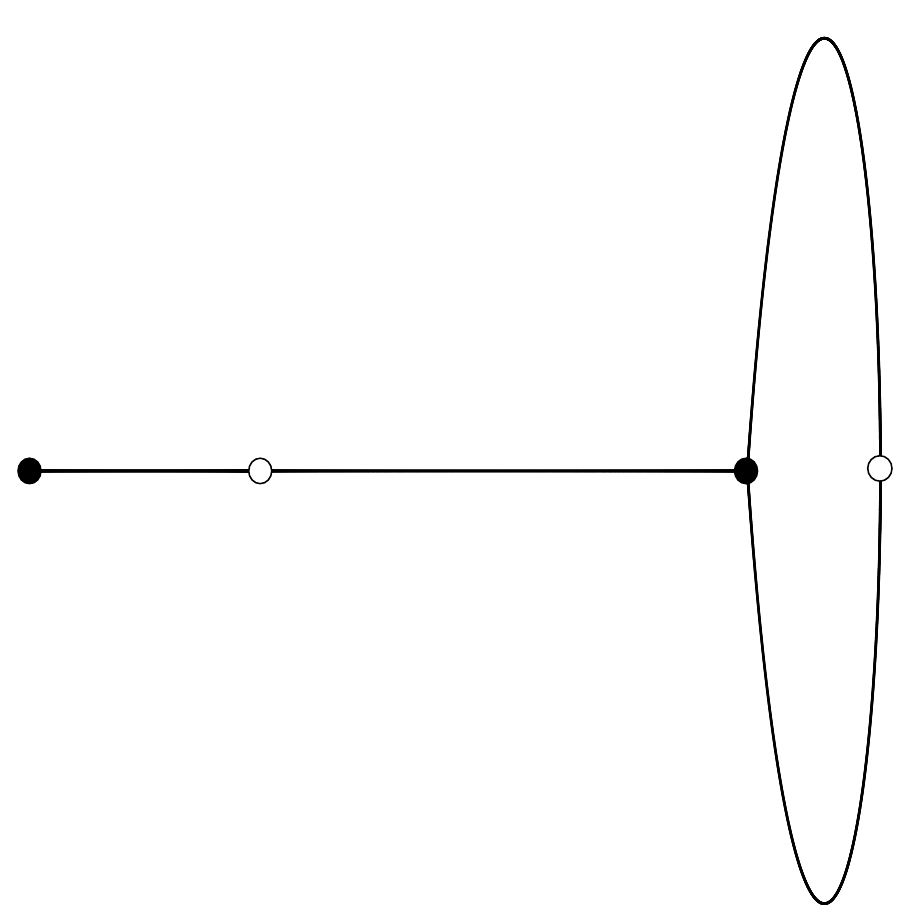}
		\caption{No.1: $\{1^13^1,2^2,1^13^1\}$, $\Gamma_0(3)$}
	\end{subfigure}
	\begin{subfigure}{6cm}
		\centering
		\includegraphics[width=5cm]{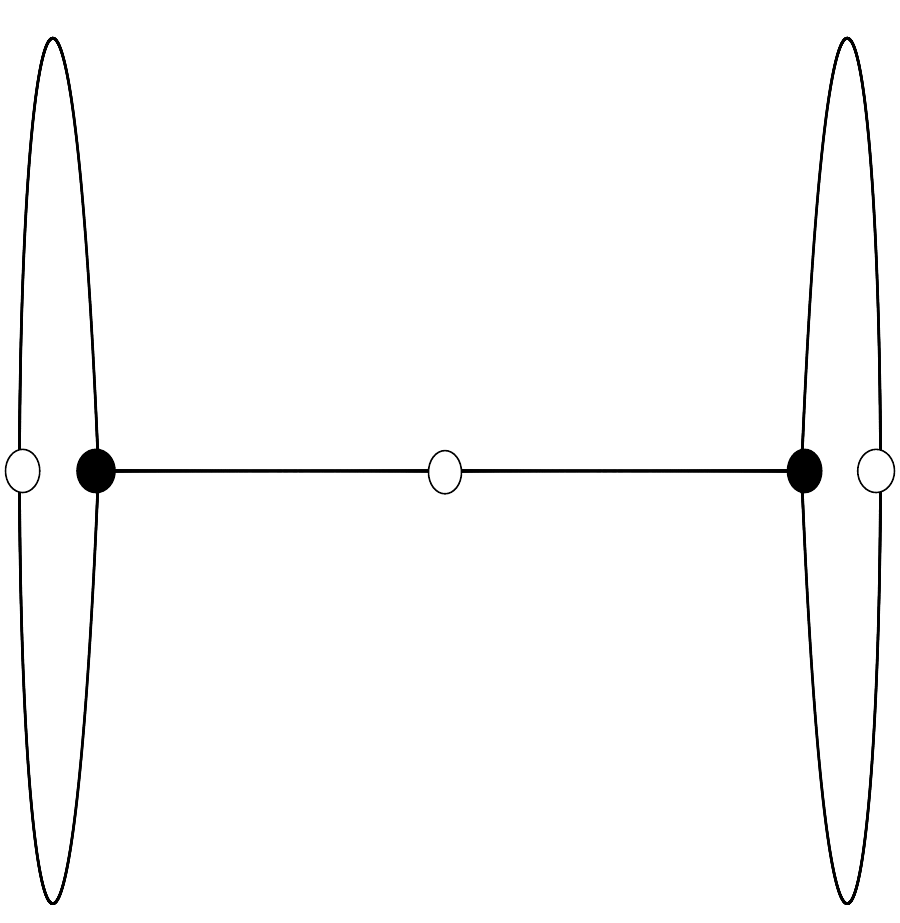}
		\caption{No.2, 3, 4: $\{3^2,2^3,1^24^1\}$, $\Gamma_0(4)$}
	\end{subfigure}
	\begin{subfigure}{6cm}
		\centering
		\includegraphics[width=5cm]{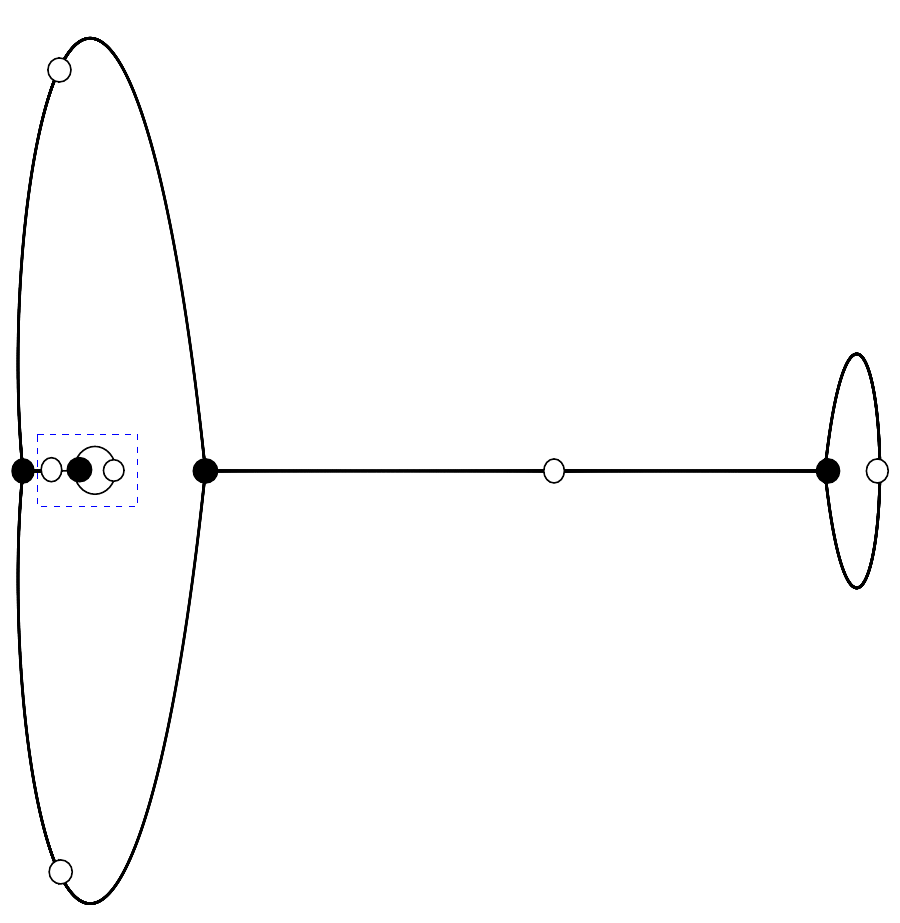}
		\caption{No.5, 6: $\{3^4,2^6,1^25^2\}$, $\Gamma_1(5)$}
	\end{subfigure}
	\begin{subfigure}{6cm}
		\centering
		\includegraphics[width=5cm]{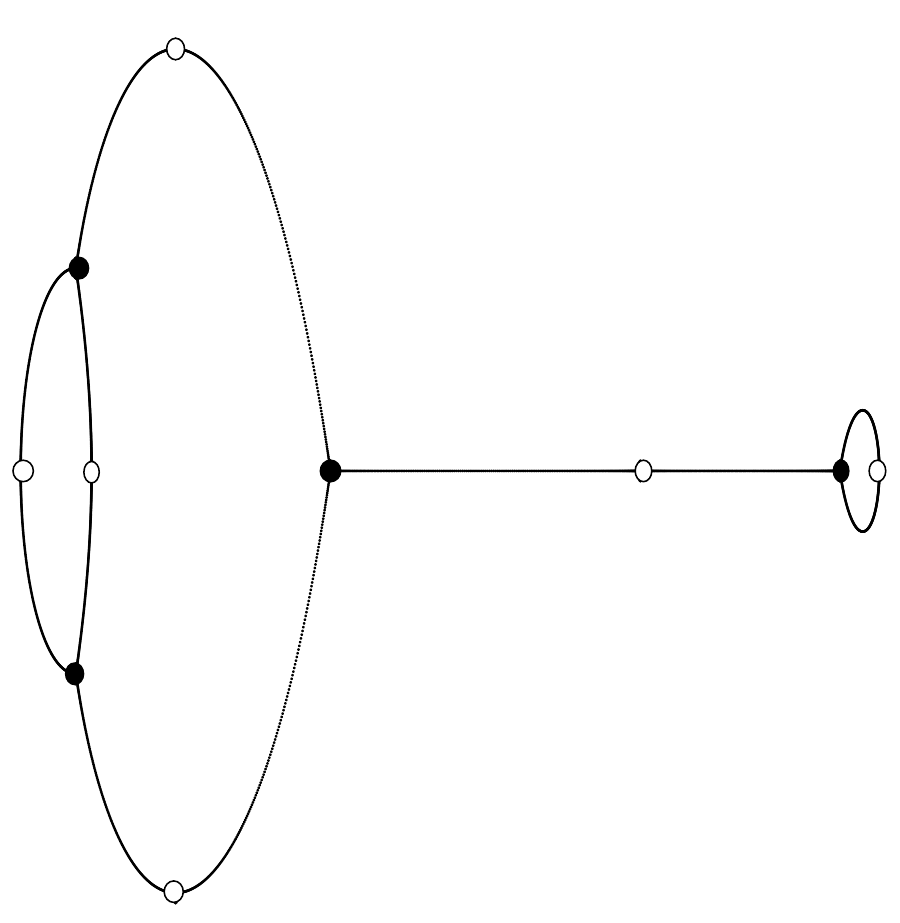}
		\caption{No.7, 8, 9, 10: $\{3^4,2^6,1^12^13^16^1\}$, $\Gamma_0(6)$}
	\end{subfigure}
	\begin{subfigure}{6cm}
		\centering
		\includegraphics[width=5cm]{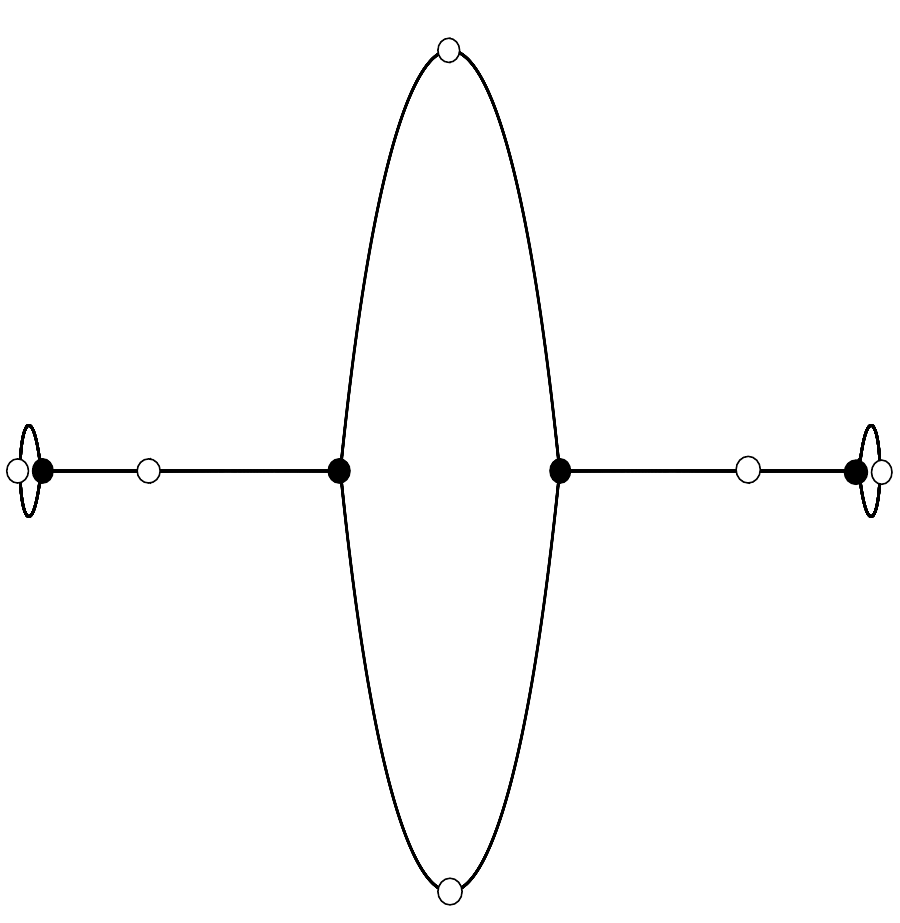}
		\caption{No.13, 15: $\{3^4,2^6,1^22^18^1\}$, $\Gamma_0(8)$}
	\end{subfigure}
	\begin{subfigure}{6cm}
		\centering
		\includegraphics[width=5cm]{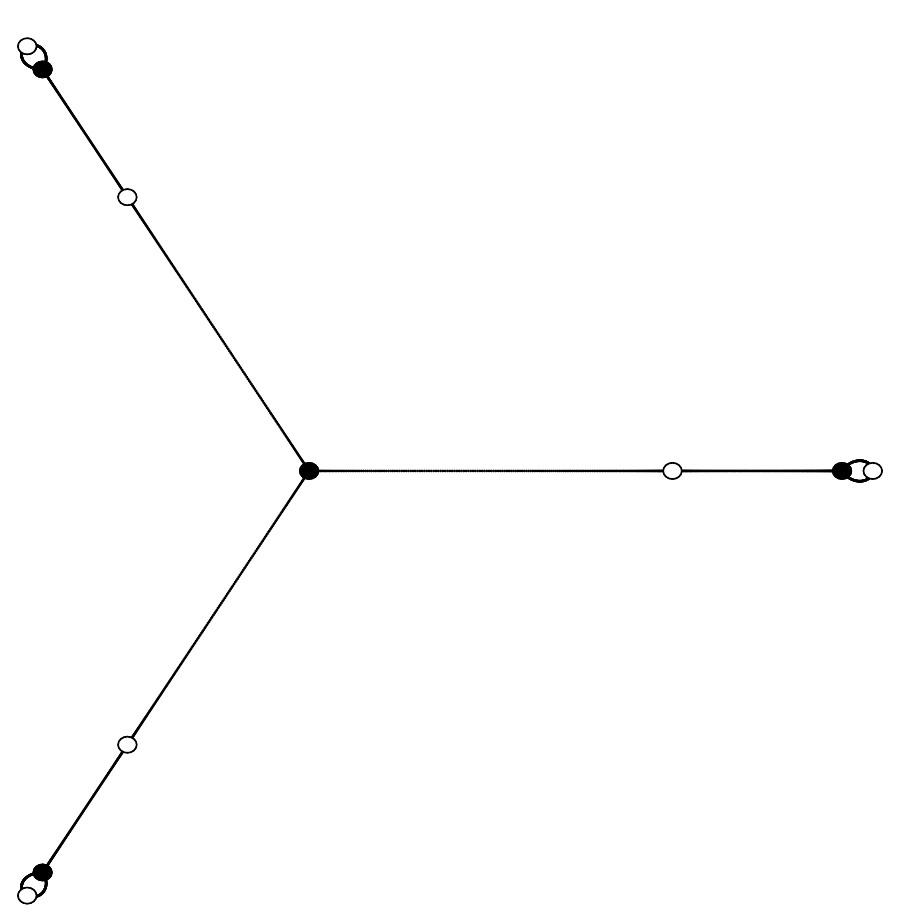}
		\caption{No.16: $\{3^4,2^6,1^39^1\}$, $\Gamma_0(9)$}
	\end{subfigure}
    \caption{The dessins for reflexive polygons with maximally tempered coefficients, their passports and the corresponding congruence subgroups. The missing of the two families (No.11, 12 and No.14) will be further discussed at the end of this subsection. Notice that in (a), the ramification is different from the form $\left\{3^V,2^E,n_1^{a_1}\dots n_k^{a_k}\right\}$, and this will be related to the discussions in \S\ref{ellpencils}.}\label{dessinref}
\end{figure}

\paragraph{Dessins d'Enfants} Given the elliptic curves in Table \ref{refelliptic}, we can then compute their $j$-invariants as in Table \ref{jinvs}.
\begin{table}[h]
\centering
\renewcommand{\arraystretch}{1.5}
\begin{tabular}{|c||c|c|c|c|}
\hline
 Polygon(s) & No.1 & No.2, 3, 4 & No.5, 6 & No.7, 8, 9, 10 \\ \hline
 $j(k)$ & $\frac{(k-18)^3(k+6)}{k-21}$ & $\frac{(k^2-8k-32)^3}{k^2-8k-48}$ & $\frac{(k^4-40k^2-120k-80)^3}{(k+3)^5(k^2-5k-25)}$ & $\frac{k^3(k^3-24k-48)^3}{(k-6)(k+2)^3(k+3)^2}$ \\ \hline\hline
 Polygon(s) & No.11, 12 & No.13, 15 & No.14 & No.16 \\ \hline
 $j(k)$ & $\frac{(k^4-16k^2-24k+16)^3}{(k+1)^2(k^3+k^2-18k-43)}$ & $\frac{(k^4-16k^2+16)^3}{k^2(k^2-16)}$ & $\frac{(k^4-8k^2-24k+16)^3}{k^4+k^3-8k^2-36k-11}$ & $\frac{k^3(k^3-24)^3}{k^3-27}$ \\ \hline
\end{tabular}
\caption{The $j$-invariants for the elliptic curves.}\label{jinvs}
\end{table}
Notice that in terms of the $k$ parameter, this is a map $j:\mathbb{P}^1\rightarrow\mathbb{P}^1,k\mapsto j(k)$. In particular, the preimage $\mathbb{P}^1\cong S^2$ is the space of $k$, and hence parametrizes the Mahler flow.
We will discuss this in more details in Appendix \ref{taubtaur}.
By further checking $j(k)/1728$, we find that all of them are Belyi (with two of the families requring a further map on $k$ which we shall discuss shortly). Therefore, we can plot the corresponding dessins as in Figure \ref{dessinref} based on the \texttt{Mathematica} package from \cite{Goins}.

To make the context more self-contained, let us explain what the information below each dessin means in Figure \ref{dessinref}. First, each dessin in Figure \ref{dessinref} is associated with a congruence subgroup. This is because the dessin can be viewed as the coset graph of the corresponding group as reviewed in \S\ref{sketch}. Now, one can construct a modular surface associated to each torsion-free, genus zero congruence subgroup \cite{shioda1972elliptic,He:2012kw,top2007explicit}. This can be done by extending the action of $\Gamma$ on the upper half plane $\mathfrak{h}$ to an automorphism
\begin{equation}
    (\mathfrak{h},\mathbb{C})\ni(\tau,z)\mapsto\left(\gamma\tau,\frac{z+m\tau+n}{c\tau+d}\right),
\end{equation}
where $\gamma\in\Gamma$ and $m,n\in\mathbb{Z}$. The modular surface then $\mathfrak{h}\times\mathbb{C}$ quotiented by this automorphism. This is often an elliptic fibration over the modular curve $X(\Gamma)$.

Moreover, we have also included the ramification data in Figure \ref{dessinref}. They are labelled as $\{3^V,2^E,n_1^{a_1}\dots n_k^{a_k}\}$. Here, $V$ and $E$ are the number of vertices and edges repectively. The number 3 (resp.~2) indicates that all the vertices are trivalent (resp.~each edge connects two vertices). The information of the faces that correspond to points ramified at $\infty$ are given by the cusp widths $\{n_i^{a_i}\}$ of the elliptic modular surface.

Here, the plots for the dessins are rigid in the sense that the vertices and edges are at the precise positions of $k=j^{-1}$ on $S^2\cong\mathbb{C}\sqcup\{\infty\}$ (except the part in the dashed blue box in (c) where we have to zoom in since the vertices $j^{-1}(0)$ and $j^{-1}(1)$ are too close to each other).

As we have checked the reflexive polygons case by case, we conclude that
\begin{proposition}
With maximally tempered coefficients for all 16 reflexive polygons, we have the following properties:
\begin{itemize}
    \item The Weierstrass normal form of the spectral curves $P(z,w)=0$, and hence the discriminants $\Delta$ and the $j$-invariants, are the same for the reflexive polygons as grouped in Table \ref{refelliptic}. In physics parlance, this grouping is based on specular duality.
    
    \item The $j$-invariants $j:\mathbb{P}^1\rightarrow\mathbb{P}^1,k\mapsto j(k)$ are Belyi maps. Hence, the reflexive polygons in the same group correspond to the same dessin.

    \item Except No. 11, 12 and No.14, the families of Weierstrass normal forms, depending on $k$, are elliptic surfaces.
\end{itemize}
\label{samedessin}
\end{proposition}

\begin{remark}
Different toric phases for a reflexive polygon are often not related by specular duality, but they would still lead to the same elliptic curve/dessin as these phases would only differ by the multiplicity of the interior point. Since the master space is invariant under specular duality, this hints that the corresponding elliptic curve and dessin should encode some common features of the master spaces in different phases.
\end{remark}

\begin{remark}
In Figure \ref{dessinref}, we have not included the reflexive polygons No.11, 12 and No. 14. This is because $j(k)/1728$ for them are ramified at four points (rather than three), and hence not Belyi. Although whether this would still give rise to any congruence subgroups is still not clear, we will discuss a possible way to map this to Belyi maps and certain dessins.
\end{remark}

\paragraph{Mahler measure} Although specular duals have the same elliptic curve, this does not directly imply that they should have the same Mahler measure as the Weierstrass normal form is obtained from the spectral curve under some bi-rational transformation while Mahler measure is only GL$(2,\mathbb{Z})$ invariant. Of course, we can compute the Mahler measures for reflexive polygons and likewise check case by case to show that specular duals have the same Mahler measure. Nevertheless, there is a more general proof using the Corollary 3.14.1 in \cite{Bao:2021fjd}:
\begin{lemma}
Given a pair of specular duals $a$ and $b$, suppose that the perfect matchings are mapped under $M_i^a\leftrightarrow M_i^b$. If their Newton polynomials are $P_{a,b}(z,w)=k-p_{a,b}(z,w)$, then for $|k|\geq\max\limits_{|z|=|w|=1}(|p_a|,|p_b|)$, the two Mahler measures have the series expansions
\begin{equation}
    m(P_{a,b})=\log(k)-\sum_{n=2}^\infty\frac{f_n(M_i^{a,b})}{nk^n},
\end{equation}
where $f_n$ are functions of $M_i^{a,b}$, and we have simply used $M_i^{a,b}$ to denote the weight for the corresponding perfect matching.
\end{lemma}

Now we can ``unrefine'' this by taking $M_i^{a,b}=1$. Then, we get the maximally tempered coefficients since they give the numbers of corresponding perfect matchings for the lattice points. Therefore,
\begin{proposition}
With maximally tempered coefficients, the Mahler measure is invariant under specular duality.\label{samemahler}
\end{proposition}
In particular,
\begin{corollary}
The reflexive polygons with maximally tempered coefficients have the same Mahler measure under specular duality.\label{samemahlerref}
\end{corollary}

\begin{remark}
As Proposition \ref{samemahler} is a general statement, if two non-reflexive polygons have specular dual phases, then they would also have the same Mahler measure. Notice however the maximally tempered coefficients would now also fix all the coefficients for the interior points to be the corresponding numbers of perfect matchings except the origin with coefficient $k$.
\end{remark}

\begin{remark}
As Mahler measure is $\textup{GL}(2,\mathbb{Z})$ invariant, equivalent lattice polygons which are classified up to $\textup{SL}(2,\mathbb{Z})$ transformations would have the same Mahler measure.
\end{remark}

For reference, we list the Mahler measures for reflexive polygons with maximally tempered coefficients in Table \ref{refmahler}.
\begingroup
\renewcommand{\arraystretch}{1.5}
\begin{longtable}{|c|c|} \hline
No.1 & $\log(k)-\frac{27}{k^2}-\frac{164}{k^3}-\frac{4941}{2k^4}-\frac{31752}{k^5}-\frac{479940}{k^6}-\frac{7426080}{k^7}$ \\
 & $-\frac{482173965}{4k^8}-\frac{6030521840}{3k^9}-\frac{171779570802}{5k^{10}}-\dots$ \\ \hline
No.2, 3, 4 & $\log(k)-\frac{10}{k^2}-\frac{32}{k^3}-\frac{297}{k^4}-\frac{2112}{k^5}-\frac{55720}{3k^6}-\frac{163200}{k^7}$ \\
 & $-\frac{3038665}{2k^8}-\frac{43406720}{3k^9}-\frac{141433992}{k^{10}}-\dots$ \\ \hline
No.5, 6 & $\log(k)-\frac{5}{k^2}-\frac{10}{k^3}-\frac{135}{2k^4}-\frac{312}{k^5}-\frac{5675}{3k^6}-\frac{11100}{k^7}$ \\
 & $-\frac{280175}{4k^8}-\frac{1346800}{3k^9}-\frac{2962386}{k^{10}}-\dots$ \\ \hline
No.7, 8, 9, 10 & $\log(k)-\frac{3}{k^2}-\frac{4}{k^3}-\frac{45}{2k^4}-\frac{72}{k^5}-\frac{340}{k^6}-\frac{1440}{k^7}$ \\
 & $-\frac{27405}{4k^8}-\frac{96880}{3k^9}-\frac{794178}{5k^{10}}-\dots$ \\ \hline
No.11, 12 & $\log(k)-\frac{2}{k^2}-\frac{2}{k^3}-\frac{9}{k^4}-\frac{24}{k^5}-\frac{245}{3k^6}-\frac{300}{k^7}$ \\
 & $-\frac{2065}{2k^8}-\frac{12320}{3k^9}-\frac{75852}{5k^{10}}-\dots$ \\ \hline
No.13, 15 & $\log(k)-\frac{2}{k^2}-\frac{9}{k^4}-\frac{200}{3k^6}-\frac{1225}{2k^8}-\frac{31752}{5k^{10}}-\dots$ \\
 & $=\log k-2k^{-2}{}_4F_3\left(1,1,\frac{3}{2},\frac{3}{2};2,2,2;16k^{-2}\right)$ \\ \hline
No.14 & $\log(k)-\frac{1}{k^2}-\frac{2}{k^3}-\frac{3}{2k^4}-\frac{12}{k^5}-\frac{55}{3k^6}-\frac{60}{k^7}$ \\
 & $-\frac{875}{4k^8}-\frac{1400}{3k^9}-\frac{9576}{5k^{10}}-\dots$ \\ \hline
No.16 & $\log(k)-\frac{2}{k^3}-\frac{15}{k^6}-\frac{560}{3k^9}-\dots$ \\
 & $=\log(k)-2k^{-3}{}_4F_3\left(1,1,\frac{4}{3},\frac{5}{3};2,2,2;27k^{-3}\right)$ \\ \hline
\caption{The Mahler measure up to order 10 for reflexive polygons with maximally tempered coefficients. Here, we restrict $k\geq\max\limits_{|z|=|w|=1}\{|p(z,w)|\}$. If $|k|\geq\max\limits_{|z|=|w|=1}\{|p(z,w)|\}$ (i.e., not necessarily real), then $\mathtt{m}(P)$ is just the real part of these expressions.}\label{refmahler}
\end{longtable}
\endgroup
This table can also be straightforwardly read off from the expressions for $u_0(k)$ in Table \ref{refu0} since
\begin{equation}
    u_0(k)=1+\sum_{n=2}^\infty\frac{a_n}{k^n},\quad m(P)=\log(k)-\sum_{n=2}^\infty\frac{a_n}{nk^n}\label{mandu0}
\end{equation}
for the same Newton polynomial.

\begin{remark}
Following \cite{Kenyon:2003uj}, the Mahler measures in Table \ref{refmahler} are the free energies of the corresponding dimer models per fundamental domain.
\end{remark}

\begin{remark}
When the Newton polynomials have maximally tempered coefficients, as both Mahler measure/$u_0(k)$ and the dessins are invariant under specular duality and should encode certain information of the master space, it would be natural to associate Mahler measures and dessins with each other.\label{mahlerdessin}
\end{remark}

\paragraph{Polygons No.11, 12 and No.14} In Figure \ref{dessinref}, we have not included the reflexive polygons No.11, 12 (which are in one family) and No.14. This is due to the fact that each $j(k)/1728$ is ramified at four points $\{0,1,x,\infty\}$. If we naively plot the pre-images $j^{-1}([0,1])$ as edges (and vertices), we would obtain the dessin for $\Gamma_0(9)$ for both of the families. This would not agree with the ramification data
\begin{equation}
    \text{No.11, 12: }\{3^4,2^6,1^32^17^1\},\quad\text{No.14: }\{3^4,2^6,1^48^1\}.
\end{equation}
In fact, we can also see that $|V|-|E|+|F|\neq2$ from the ramification data. The reason is that there would be another pre-image of $\infty$ in the same face for $k=\infty$ if we put the ramification data on the dessin for $\Gamma_0(9)$. This indicates that there are more ramification points besides $\{0,1,\infty\}$. Indeed, we find that each of them has an extra ramification:
\begin{align}
    &\text{No.11, 12: }\frac{1}{1728}j(-17/7)=-\frac{665492189}{5692329216},\\
    &\text{No.14: }\frac{1}{1728}j(-9/8)=\frac{35163174217}{28991029248}.
\end{align}
Therefore, they are not Belyi any more. Nevertheless, we can modify the standard transformation $k \mapsto \frac{(m+n)^{m+n}}{m^mn^n}k^m (1-k)^n$ from \cite{belyui1980galois} which takes $\{ 0, 1, \frac{m}{m+n}, \infty \}$ to $\{ 0, 1, \infty \}$ in order to eliminate the extra ramification point.
However, it is not clear any more whether the resulting dessins would correspond to any congruence subgroups. For No.11, 12, we take the map
\begin{equation}
    k\mapsto-\frac{5692329216^{5692329216}}{665492189^{665492189}5026837027^{5026837027}}k^{665492189}(1-k)^{5026837027}
\end{equation}
while for No.14, we take the map
\begin{equation}
    k\mapsto\frac{6172144969^{6172144969}28991029248^{28991029248}}{35163174217^{35163174217}}k^{28991029248}\left(1-\frac{1}{k}\right)^{-6172144969}.
\end{equation}
This would lead to dessins with a huge number of vertices, and hence we shall not plot them here.

For physicists, the exceptions of No.11, 12 and No.14 also have their incarnations in the context of string theory and quantum field theories. For instance, as shown in \cite{Closset:2021lhd}, the massless Coulomb branches of rank-1 $E_n$ 5d SCFTs, whose brane web constructions are dual to the (P)dP$_{n}$ polygons, are exactly the modular curves associated to the congruence of the dessins, except $\Tilde{E}_1$ (No.14) and $E_2$ (No.12). Indeed, the massless Coulomb branches for the theories $\Tilde{E}_1$ and $E_2$ are not modular curves due to the fact that the 5d SCFTs are special with abelian flavour symmetries. The $k$ parameter is closely related to the $U$-plane in such context. Later in \S\ref{ellpencils}, we will also see that the connections to the elliptic pencils are missing for these two families.

\subsection{Quantum Periods for del Pezzo Surfaces}\label{delpezzo}
In \cite{coates2016quantum}, the quantum periods
\begin{equation}
    G_X(\lambda)=1+\sum_{n=2}^{\infty}\frac{a_n}{n!}\lambda^n
\end{equation}
for the del Pezzo surfaces $X$ were computed. It is worth noting that the coefficients coincide with the ones in the Mahler measure for Newton polynomials with maximally tempered coefficients
\begin{equation}
    m(P)=-\log(\lambda)-\sum_{n=2}^{\infty}\frac{a_n}{n}\lambda^n
\end{equation}
according to Table \ref{GX}.
\begingroup
\renewcommand{\arraystretch}{1.5}
\begin{longtable}{|c|c|} \hline
 Reflexive polygon(s) & del Pezzo surface $X$ \\ \hline
No.1 & degree 3 \\ \hline
No.2, 3, 4 & degree 4 \\ \hline
No.5, 6 & degree 5 \\ \hline
No.7, 8, 9, 10 & degree 6 \\ \hline
No.11, 12 & degree 7 \\ \hline
No.14 & degree 8: $\mathbb{F}_1$ \\ \hline
No.13, 15 & degree 8: $\mathbb{P}^1\times\mathbb{P}^1$ \\ \hline
No.16 & degree 9: $\mathbb{P}^2$ \\ \hline
\caption{The correspondence between $m(P)$ and $G_X(\lambda)$.}\label{GX}
\end{longtable}
\endgroup
In other words, the (normalized) areas $A(\mathfrak{P})$ of the reflexive polygons are related to the degrees of the del Pezzo surfaces by\footnote{For del Pezzo surfaces of degrees 1 and 2, they can actually be related to certain reflexive polygons with minimally tempered coefficients. See Appendix \ref{minimally}.}
\begin{equation}
    A(\mathfrak{P})=12-\text{deg}(X).
\end{equation}
As the coefficients $a_n$ are the constant terms of $p(z,w)^n$, we can see that $u_0$ is the Laplace transform of $G_X$. This correspondence can also be verified by the integral representations
\begin{align}
    &m(P)=\frac{1}{(2\pi i)^2}\int_{|z|=|w|=1}\log\left(\lambda^{-1}-p(z,w)\right)\frac{\text{d}z}{z}\frac{\text{d}w}{w},\\
    &G_X(\lambda)=\frac{1}{(2\pi i)^2}\int_{|z|=|w|=1}\exp(\lambda p(z,w))\frac{\text{d}z}{z}\frac{\text{d}w}{w}.
\end{align}
This is in fact a reflection of the mirror symmetry \cite{Strominger:1996it} that relates the del Pezzo surfaces and the Landau-Ginzburg models where the Newton polynomials play the roles as the superpotentials\footnote{See also \cite{fei2021mahler} for some relevant discussions.}. The study of $G_X(\lambda)$ belongs to the broad picture of Golyshev's programme \cite{golyshev2005classification} that aims to translate problems in algebraic geometry into problems in modular forms and differential equations via mirror symmetry.

\subsection{Hauptmoduln and the $k$ parameter}\label{hauptmoduln}
In this subsection, we shall give more clues on the connection between Mahler measure and dessins, as well as to congruence groups. Let us consider the modular expansion of Mahler measure and illustrate this with a few examples.

\paragraph{Example 1: No.15} As reviewed in \S\ref{modularmahler}, the Mahler measure for $P=k-z-z^{-1}-w-w^{-1}$ reads
\begin{equation}
    m(P)=\frac{16\text{Im}\tau}{\pi^2}\sum_{\substack{n_1,n_2\in\mathbb{Z}\\(n_1,n_2)\neq(0,0)}}\frac{\chi_{-4}(n_1)}{(n_1+4n_2\tau)^2(n_1+4n_2\Bar{\tau})}.
\end{equation}
Of particular interest here would be the parameter $k$, where \cite{samart2014mahler}
\begin{equation}
    k^2=\frac{\eta^{24}(2\tau)}{\eta^8(\tau)\eta^{16}(4\tau)}=q^{-1}+8+20q-62q^3+216q^5-\dots,
\end{equation}
with $\eta(\tau)$ being the Dedekind eta function. This is a Hauptmodul\footnote{The Hauptmoduln for some $\Gamma_0(N)$ can also be found for example in \cite{beneish2015traces}.} for $\Gamma_0(4)$. In particular, the congruence subgroup associated to the dessin in this case is $\Gamma_0(8)$, which is a subgroup of $\Gamma_0(4)$.

\paragraph{Example 2: No.16} The Newton polynomial is $P=k-z-w-z^{-1}w^{-1}$. One can compute that
\begin{equation}
    u_0(k)={}_2F_1\left(\frac{1}{3},\frac{2}{3};1;\frac{27}{k^3}\right).
\end{equation}
For convenience, we write $\mu\equiv1/k^3$. Then we have \cite{villegas1999modular}
\begin{equation}
    \begin{split}
        &u_0=1+6\sum_{n=1}^{\infty}\sum_{d|n}\chi_{-3}(d)q^n,\quad e=1-9\sum_{n=1}^{\infty}\sum_{d|n}d^2\chi_{-3}(d)q^n,\\
        &\mu=\frac{1}{27}\left(1-\frac{e}{c^3}\right)=q-15q^2+171q^3-1679q^4+\dots,
    \end{split}
\end{equation}
where $\chi_{-3}(n)=0,1,-1$ when $n\equiv0,1,2~(\text{mod }3)$. The Mahler measure is
\begin{equation}
    m(P)=\frac{81\sqrt{3}\text{Im}\tau}{4\pi^2}\sum_{\substack{n_1,n_2\in\mathbb{Z}\\(n_1,n_2)\neq(0,0)}}\frac{\chi_{-3}(n_1)}{(n_1+3n_2\tau)^2(n_1+3n_2\Bar{\tau})}.
\end{equation}
Moreover, we have \cite{samart2014mahler}
\begin{equation}
    k^3=1/\mu=27+\left(\frac{\eta(\tau)}{\eta(3\tau)}\right)^{12}=27+\frac{1}{q}-12+54q-76q^2-243q^3+1188q^4-1384q^5+\dots
\end{equation}
This is a Hauptmodul for $\Gamma_0(3)$. In particular, the congruence subgroup associated to the dessin in this case is $\Gamma_0(9)$, which is a subgroup of $\Gamma_0(3)$.

\paragraph{Example 3: No. 5, 6} The Newton polynomials are $P=k-z-z^{-1}w^2-z^{-1}w^{-1}-w^{-1}-2w-3z^{-1}w-3z^{-1}$ for No.5 and $P=k-z-w-z^{-1}w-z^{-1}w^{-1}-zw^{-1}-2z^{-1}-2w^{-1}$ for No.6. This has actually been computed in \cite{zagier2009integral,Stienstra:2005wy}:
\begin{equation}
    \begin{split}
        &u_0=1+\frac{1}{2}\sum_{n=1}^{\infty}((3-\text{i})\chi(n)+(3+\text{i})\overline{\chi(n)})\frac{q^n}{1-q^n},\\
        &e=1+\frac{1}{2}\sum_{n=1}^{\infty}((2-\text{i})\chi(n)+(2+\text{i})\overline{\chi(n)})\frac{n^2q^n}{1-q^n},\\
        &m(P)=-2\pi\text{i}\tau-\frac{1}{2}\sum_{n=1}^{\infty}\sum_{d|n}((2-\text{i})\chi(d)+(2+\text{i})\overline{\chi(d)})nq^n,
    \end{split}
\end{equation}
where $\chi(n)=\text{i}^l$ when $n\equiv2^l~(\text{mod }5)$. Moreover, we have
\begin{equation}
    (k-3)^{-1}=q\prod_{n=1}^{\infty}(1-q^n)^{5\left(\frac{n}{5}\right)}=3+q-5q^2+15q^3-30q^4+40q^5-\dots,
\end{equation}
where $\left(\frac{n}{5}\right)=(-1)^l$ when $n\equiv2^l~(\text{mod }5)$. This is a Hauptmodul for $\Gamma_1(5)$. In particular, the congruence subgroup associated to the dessin in this case is $\Gamma_1(5)$.

\paragraph{Example 4: No.7, 8, 9, 10} This has actually been computed in \cite{zagier2009integral,Stienstra:2005wy}:
\begin{equation}
    \begin{split}
        &u_0=\frac{\eta(2\tau)\eta^6(3\tau)}{\eta^2(\tau)\eta^3(6\tau)},\\
        &e=1+\sum_{n=1}^{\infty}(-1)^n\chi_{-3}(n)\frac{n^2q^n}{1-q^n},\\
        &m(P)=-2\pi\text{i}\tau-\sum_{n=1}^{\infty}\sum_{d|n}(-1)^d\chi_{-3}(d)nq^n,
    \end{split}
\end{equation}
where $\chi_{-3}(n)$ is the same as in Example 2. Moreover, we have
\begin{equation}
    k-2=\frac{\eta^3(2\tau)\eta^9(3\tau)}{\eta^3(\tau)\eta^9(6\tau)}=\frac{1}{q}+3+6q+4q^2-3q^3-12q^4-8q^5+\dots
\end{equation}
This is a Hauptmodul for $\Gamma_0(6)$. In particular, the congruence subgroup associated to the dessin in this case is $\Gamma_0(6)$.

As we can see, the $k$ parameter is closely related to the Hauptmodul of certain congruence subgroup\footnote{Therefore, the Hauptmoduln, and hence the meromorphic functions on modular curves, should physically be related to the (sizes of) gas phases for dimer models and the Mahler flows \cite{Bao:2021fjd,Kenyon:2003uj}.}. We may also conjecture that
\begin{conjecture}
Let $\Gamma^a$ be the congruence subgroup associated to the dessin for the reflexive polygons in Figure \ref{dessinref} (with maximally tempered coefficients). Then $k^n$ is a Hauptmodul for some congruence subgroup $\Gamma^b$, and $\Gamma^a\leq\Gamma^b$. Moreover, $\Gamma^b$ is the monodromy group of the corresponding Picard-Fuchs equation. \label{hauptmodulconj1}
\end{conjecture}
We may even give a stronger conjecture.
\begin{conjecture}
If $\Gamma^a=\Gamma_{l_1}(r_1)$ and $\Gamma^b=\Gamma_{l_2}(r_2)$ (where $l_{1,2}=0,1$), then $l_1=l_2$ and $r_1=|n_2|r_2$.\label{hauptmodulconj2}
\end{conjecture}

From the above examples, we can see that there are two remaining families in Figure \ref{dessinref} where the statements are not proven, namely, No.1 and No.2, 3, 4. At the current stage, we have not found a good way to obtain the explicit Hauptmoduln for them. Nevertheless, we note that for the known examples here, the congruence subgroups are all of index 12. In other words, the sums of their cusp widths are always 12. For No.1 and No.2, 3, 4, however, the indices of the congruence subgroups are 4 and 6 respectively.

Here, we are focusing on the maximally tempered coefficients. Mathematically, we would also wonder whether the $k$ parameters could be related to Hauptmoduln for certain congruence subgroups for any coefficients. In Appendix \ref{minimally}, we give different types of examples for minimally tempered coefficients.

\subsection{Elliptic Pencils}\label{ellpencils}
Another intriguing observation is the connections to the Beauville pencils \cite{beauville1982familles} (see also \cite{zagier2009integral}). This classifies the stable families of elliptic curves over $\mathbb{P}^1$ with exactly four singular fibres\footnote{The minimal number of singular fibres of a stable family of elliptic curves is four.}. In particular, the fibres of a Beauville pencil correspond to points on certain modular curves, and hence each family has an associated the congruence subgroup. Among the six cases whose table in \cite{beauville1982familles,zagier2009integral} is reproduced here in Table \ref{Beauville}, four of them (III$\sim$VI) are exactly the reflexive polygons with maximally tempered coefficients.
\begingroup
\renewcommand{\arraystretch}{1.5}
\begin{longtable}{|c|c|c|} \hline
 Case & Equation of the family & Group \\ \hline
I & $X^3+Y^3+Z^3+tXYZ=0$ & $\Gamma(3)$ \\ \hline
II & $X(X^2+Z^2+2ZY)+tZ(X^2-Y^2)=0$ & $\Gamma_1(4)\cap\Gamma(2)$ \\ \hline
III & $X(X-Z)(Y-Z)+tZY(X-Y)=0$ & $\Gamma_1(5)$ \\ \hline
IV & $(X+Y)(Y+Z)(Z+X)+tXYZ=0$ & $\Gamma_1(6)$ \\ \hline
V & $(X+Y)(XY-Z^2)+tXYZ=0$ & $\Gamma_0(8)\cap\Gamma_1(4)$ \\ \hline
VI & $X^2Y+Y^2Z+Z^2X+tXYZ=0$ & $\Gamma_0(9)\cap\Gamma_1(3)$ \\ \hline
\caption{The Beauville pencils.}\label{Beauville}
\end{longtable}
\endgroup
More explicitly, we have
\begin{itemize}
    \item Case III: This corresponds to No.6 under the substitution
    \[(X,Y,Z)\mapsto(X+Y,Y,-Z),t\mapsto-t.\]
    
    \item Case IV: This corresponds to No.10.

    \item Case V: This corresponds to No.15 under the substitution
    \[(X,Y,Z)\mapsto(X,Y,iZ),t\mapsto-it.\]

    \item  Case VI: This corresponds to No.16.
\end{itemize}
One may also check that the $j$-invariants obtained therefrom (as listed in the bottom table on page 13 in \cite{zagier2009integral}) agrees with the results in Table \ref{jinvs}. It is then not hard to see that
\begin{proposition}
    The congruence subgroups for the Beauville pencils are subgroups of the congruence subgroups associated to the dessins for the corresponding reflexive polygons. In this sense, the Beauville pencils are in one-to-one correspondence with the families of the reflexive polygons with maximally or minimally tempered coefficients whose dessins are of index 12. In particular, the elliptic surfaces are modular.\label{Beauvilleprop}
\end{proposition}
For this proposition to hold, we also need to include families I and II. In fact, we notice that the singular fibres for these two families have types other than $I_n$ (see Table \ref{singfibs} below), which indicates that the corresponding elliptic surfaces are not modular. More concretely, for Case I (resp.~II), at the singular fibre $IV^*$ (resp.~$I^*$), the discriminant has a zero of multiplicity 8 (resp.~7), and the $j$-function has a zero (resp.~a pole of order 1). On the other hand, at a singular fibre $I_n$, the discriminant has a zero of multiplicity $n$, and the $j$-function has a pole of order $n$. For Cases I and II, they are actually related to the minimally tempered cases and are discussed in Appendix \ref{minimally}.

As we can see, every reflexive polygons with dessin of index 12 is naturally related to some Beauville pencil. For the remaining two dessins (with different indices), they can also be related to some elliptic pencils. From \cite{SchmicklerHirzebruch1985ElliptischeF}, one can find that the singular fibre configurations (see for example Table \ref{singfibs} below) are the elliptic pencils of the following congruence subgroups:
\begin{itemize}
    \item No.1: $\Gamma_1(3)$;
    \item No.2, 3, 4: $\Gamma_1(4)$.
\end{itemize}
Again, we can see that they are subgroups of the congruence subgroups associated to the dessins. Therefore,
\begin{proposition}
    The congruence subgroup for the elliptic pencil associated to a reflexive polygon with maximally tempered coefficients is a subgroup of the congruence subgroup associated to the dessin. Hence, assuming Conjecture \ref{hauptmodulconj1}, the congruence subgroup for the elliptic pencil is a subgroup of the monodromy group of the Picard-Fuchs equation.
 \label{ellpenprop2}
\end{proposition}

Let us say something more about the modularity of the elliptic pencils. For each case, we have a family of isomorphic elliptic curves parametrized by $t$. Now, being modular indicates the property that all the elliptic curves in a family contain a specific subgroup identified under the isomorphisms. Let us illustrate this with an example.
\begin{example}
    Take Case IV in Table \ref{Beauville} with the associated congruence subgroup $\Gamma_1(6)$, which corresponds to the polygon No.10. For any $t$, the elliptic curve would always pass through the following six points:
    \begin{align}
        &P_1=[1:0:0],\quad P_2=[0:1:0],\quad P_3=[0:0:1],\nonumber\\
        &P_4=[0:1:-1],\quad P_5=[1:0:-1],\quad P_6=[1:-1:0].
    \end{align}
    The standard group law on the elliptic curve gives rise to an abelian group, and these six points in fact form a common subgroup $\mathbb{Z}_6$ in this family. The group elements are given as follows:
    \begin{equation}
        P_4=0,\quad P_2=P_2,\quad P_6=2P_2,\quad P_1=3P_2,\quad P_5=4P_2,\quad P_3=5P_2.
    \end{equation}
    In other words, $P_4$ is the identity and $P_2$ is the generator (i.e., $6P_2=0$). Let us verify this claim. For any $t$, the elliptic curve would intersect the following seven lines at the above six points:
    \begin{align}
        &X=0:\quad\quad\quad\quad\quad P_2,P_3,P_4;\\
        &Y=0:\quad\quad\quad\quad\quad P_1,P_3,P_5;\\
        &Z=0:\quad\quad\quad\quad\quad P_1,P_2,P_6;\\
        &Y+Z=0:\quad\quad\quad P_1,P_1,P_6;\\
        &X+Z=0:\quad\quad\quad P_2,P_2,P_5;\\
        &X+Y=0:\quad\quad\quad P_3,P_3,P_6;\\
        &X+Y+Z=0:\quad P_4,P_5,P_6.
    \end{align}
    Here, repeated points indicate the multiplicities of the intersections. Indeed, the group elements satisfy the relations from these lines. For instance, for $X=0$, we have
    \begin{equation}
        P_2+P_3+P_4=P_2+5P_2+0=6P_2=0.
    \end{equation}
    The others can also be checked in a straightforward manner.
\end{example}

Recall that No.11, 12 and No.14 are exceptional as $j(k)/1728$ are not Belyi. In fact, we do not find any elliptic pencils associated to their ramification data and singular fibre configurations.

\section{Further Connections to String/F-Theory}\label{Fthy}
In this section, we shall briefly discuss some relations with F-theory compactification, especially relating the dessins and 7-branes. If we consider a sigma model whose target space is one of the non-compact CY 3-folds from the reflexive polygons, then its mirror is Landau-Ginzburg theory with the $W$-plane $W=P(z,w)$. In particular, the BPS states from D-branes wrapping compact cycles can be studied via some F-theory background \cite{Hori:2000ck}.

We recall that given an elliptic fibration over some complex base $B$ with fibre $y^2=x^3+f(v)x+g(v)$ and $v\in B$, the F-theory compactified on it is equivalent to Type IIB compactification on $B$ with complexified coupling $\tau$. This coupling $\tau$, which serves as the complex structure of the elliptic fibre, can be  exactly identified as $\tau=\frac{1}{2\pi i}\frac{u_1}{u_0}$ in our modular Mahler measure discussions, and is defined up to SL$(2,\mathbb{Z})$ transformations.

As mentioned before, the elliptic curve becomes singular and the fibre degenerates when the discriminant $\Delta=4f^3+27g^2$ vanishes. These are the positions where $(p,q)$ 7-branes are placed since $\tau$ is transformed by SL$(2,\mathbb{Z})$ transformations under the monodromies around 7-branes in Type IIB or in F-theory. The detailed background on the 7-branes are not so important here, and interested readers are referred to for example \cite{Heckman:2018jxk} for a nice review on relevant topics. For more mathematically oriented audience, it suffices to know that when one constructs quantum field theories via the so-called F-theory on elliptically fibred CYs, the 7-branes are put exactly on the singular fibres. They fill seven spatial directions (and one temporal direction) in the 10-dimensional spacetime (the extra two dimensions in F-theory are the elliptic fibre). In other words, when we factorize the discriminant as $\Delta=\prod\limits_a\Delta_a$, each 7-brane is essentially a factor $\Delta_a$. In physical theories, the 7-branes have many key features. For instance, each 7-brane has a branch cut, which gives us a monodromy when one passes through it. From this one can obtain the brane creations and annihilations under Hanany-Witten transitions \cite{Hanany:1996ie} that are crucial in the brane set-up of the theories. Different 7-branes/singular fibres would give rise to different gauge algebras under which the matters transform. Therefore, the rules of possible gauge algebras as given in \cite{Bershadsky:1996nh,Katz:1996xe,Aspinwall:2000kf} are exactly based on Kodaira's classification of singular fibre types.

Here, let us consider the surface that corresponds to a toric diagram, which defines the CY singularity, or we can think of the geometry as a double fibration over the $W$-plane with a $\mathbb{C}^*$ fibre and a punctured Riemann surface $W=P(z,w)$. In particular, the surface is now over the base $\mathbb{P}^1$ parametrized by the $k$ parameter with fibre $P(z,w)$. For reflexive polygons (with maximally temperd coefficients)\footnote{We also list the types of the singular fibres for the minimally tempered coefficients in Appendix \ref{minimally}.} in this paper, the singular fibres are listed in Table \ref{singfibs}. As we can see, this agrees with the ramification data in \S\ref{dessinsMahler}. The fibre types at $k=\infty$ also coincide with the degrees of the del Pezzo surfaces in Table \ref{GX}.
\begingroup
\renewcommand{\arraystretch}{1.5}
\begin{longtable}{|c|c|} \hline
 Reflexive polygon(s) & Singular fibres \\ \hline
No.1 & $I_3,I_1,IV^*$ \\ \hline
No.2, 3, 4 & $I_4,I_1,I_1^*$ \\ \hline
No.5, 6 & $I_5,I_5,I_1,I_1$ \\ \hline
No.7, 8, 9, 10 & $I_6,I_3,I_2,I_1$\\ \hline
No.11, 12 & $I_7,I_2,I_1,I_1,I_1$ \\ \hline
No.14 & $I_8,I_1,I_1,I_1,I_1$ \\ \hline
No.13, 15 & $I_8,I_2,I_1,I_1$ \\ \hline
No.16 & $I_9,I_1,I_1,I_1$ \\ \hline
\caption{The singular fibre types determined from $f,g$ and $\Delta$ associated to the reflexive polygons. The fibre at $k=\infty$ is listed first in each row.}\label{singfibs}
\end{longtable}
\endgroup

From Table \ref{refelliptic}, we know that in our cases $f(k)$ and $g(k)$ are always of degrees 4 and 6 respectively. Therefore, one would expect $\Delta$ to be of degree 12. This would agree with the requirement of 12 7-branes in physics. However, it is possible for $\Delta$ to have degree less than 12. The reasons are that $4f^3+27g^2$ may have cancellations of terms. Nevertheless, as we shall now discuss, we are still able to recover the 12 7-branes, and we can actually put them on the dessin.

As the fibre degenerates at the $n$ zeros of $\Delta$ (counted with multiplicity), there must be $n$ 7-branes associated to them. It turns out that the remaining $(12-n)$ 7-branes are compensated by $j\rightarrow\infty$ at the tropical limit, that is, $k\rightarrow\infty$. Indeed, by checking the degree of the numerator minus the degree of the denominator of $j$ in Table \ref{jinvs}, we find that they are precisely equal to 12 minus the degree of $\Delta$. This actually makes sense since we are now considering the compact $\mathbb{P}^1$ as the space of $k$. Therefore, we should also take the singular curve at $k$ tropical into account, which is just a usual point on the compact sphere.

As the corresponding dessin is parametrized by $\beta=j/1728$, one may consider to associate the 7-branes to the faces of the dessin. However, some 7-branes could still not correspond to the faces (both internal and external), i.e., $j\rightarrow\infty$. This is because the numerator and denominator of $j$ may have some factors being cancelled. Suppose the $j$-invariant is
\begin{equation}
    j=-\frac{2\times(24f)^3}{\Delta}=\frac{(k-k_*)^{n_1}f_1(k)}{(k-k_*)^{n_2}f_2(k)},
\end{equation}
where $n_1>n_2$ and $f_{1,2}$ do not have any $(k-k_*)$ factor. As a result, $k=k_*$ is a zero of $\Delta$ which makes the curve singular, but this information is not encoded by $j\rightarrow\infty$ since such factors all get cancelled in the denominator. Nevertheless, as $n_1>n_2$, we find that such 7-branes now correspond to a black node (pre-image of $j=0$) in the dessin.

Notice that we have not considered the possibility of $n_1=n_2$. If so, then such number of 7-branes would not correspond to a face or a black vertex in the dessin. We shall then write\footnote{Notice that we do not have any further restrictions on $f_a$ and $g_b$, so they could still have common factors. However, \eqref{fagb} suffices to complete our argument as we only need to know whether 7-branes could be associated to places other than faces and black vertices.}
\begin{equation}
    f=(k-k_*)^{2n}f_a,\quad g=(k-k_*)^{3n}g_b,\label{fagb}
\end{equation}
where the subscripts $a,b$ indicate the degree of $f_a$ and $g_b$. This is because now the $j$-invariant looks like
\begin{equation}
    j=-2\times24^3\times\frac{(k-k_*)^{6n}f_a^3}{4(k-k_*)^{6n}f_a^3+27(k-k_*)^{6n}g_b^2}.
\end{equation}
Since the degrees of $f$ and $g$ are 4 and 6 respectively, $n$ can only be 1 or 2.

Let us first consider the case $n=2$. Then $a=b=0$. In other words, $f^3\propto g^2$. In this case, the $j$-invariant is trivially a constant, and the dessin is empty. Equivalently, we can think of it as the external face where all the 7-branes live being the whole sphere with no other elements for the dessin.

If $n=1$, then we can write the elliptic curve as
\begin{equation}
    \frac{y^2}{(k-k_*)^3}=\frac{x^3}{(k-k_*)^3}+\frac{f_2x}{(k-k_*)}+g_3.
\end{equation}
Under the redefinition $x/(k-k_*)\rightarrow x$ and $y/(k-k_*)^{3/2}\rightarrow y$, we get the Weierstrass normal form
\begin{equation}
    y^2=x^3+f_2x+g_3,
\end{equation}
where $f_2$ and $g_3$ are of degrees 2 and 3. Hence, no matter what value $k_*$ is, we would only get the same curve, and we are only left with six 7-branes.

We have therefore shown that
\begin{proposition}
On the dessin, all the faces (including both internal and external) and some of the black vertices (the pre-images $j=0$) correspond to $7$-branes. A black vertex at $k=k_*$ is associated to $7$-branes if and only if $j(k_*)=0$ and $\Delta(k_*)=0$.\label{branefaceblack}
\end{proposition}

\begin{example}
Let us illustrate this with an example whose $7$-branes are associated to both faces and a black vertex. The dessin for the reflexive polygon No.1 is given in Figure \ref{dessinref}(a). Moreover,
\begin{equation}
    \Delta=(k+6)^8(k-21),\quad j=\frac{(k-18)^3(k+6)}{k-21}.
\end{equation}
Hence, there is a $7$-brane located at the centre $k=21$ of the internal face. Moreover, since there is a zero of order $8$ for $\Delta$ at $k=-6$, this would give eight $7$-branes on top of each other. As $j$ is also zero in this case, we find that the eight $7$-branes correspond to the leftmost black vertex in the dessin. So far, we have only found nine $7$-branes. The remaining three are placed at the tropical $k$ in the external face on the sphere. Indeed, we have $j\rightarrow k^3$ when $k\rightarrow\infty$.
\end{example}

It is worth noting that when $j$ diverges at say $k=k_*$, near this point we have $j\sim\frac{1}{k-k_*}$. This yields $\tau\sim\frac{1}{2\pi i}\log(k-k_*)$. When $k\rightarrow k_*$, we get $\tau\rightarrow i\infty$. As $\tau=\frac{\theta}{2\pi}+\frac{i}{g_{\text{IIB}}}$, we have $g_{\text{IIB}}\rightarrow0$. Notice that this weak coupling regime is only local due to the SL$(2,\mathbb{Z})$ transformation. In the special case when $f^3\propto g^2$, $j$ becomes a constant. In particular, when $f^3/g^2$ is $\frac{-3}{4^{1/3}}$, we have a global weak coupling \cite{Sen:1996vd,Sen:1997gv}.

\paragraph{Brane monodromy and dessin monodromy} The non-trivial effect of passing the branch cut of a $(p,q)$ 7-brane is often encoded by the monodromy matrix $M_{p,q}\in\text{SL}(2,\mathbb{Z})$ \cite{Gaberdiel:1998mv}. In fact, we can relate the monodromy group $G$ of the dessin generated by $(\sigma_0,\sigma_1,\sigma_\infty)$ to the monodromies of the 7-branes.

The general strategy is as follows. First, we choose a reference point on the dessin, just like what one does for 7-branes. As the monodromy for a 7-brane is analyzed by a loop going around the branch cut connecting the brane and the reference point, we also go along the loops on the dessin surrounding the reference point and the internal faces/black vertices where the 7-branes are. Then these loops would correspond to some permutations $\sigma^i\in G$ which can be obtained from the generators $(\sigma_0,\sigma_1)$. Finally, we can determine the permutation for the external face, namely the tropical limit $k\rightarrow\infty$, using $\prod\sigma^i=1$ as $\prod M_{p,q}=1$. Notice that this identity also guarantees that the permutation for the external face must be an element of $G$.

\begin{example}
Let us illustrate this again with the reflexive polygon No.1. In Figure \ref{monodromyex}, we label the edges and plot the monodromies explicitly on the dessin.
\begin{figure}[h]
    \centering
    \includegraphics[width=6cm]{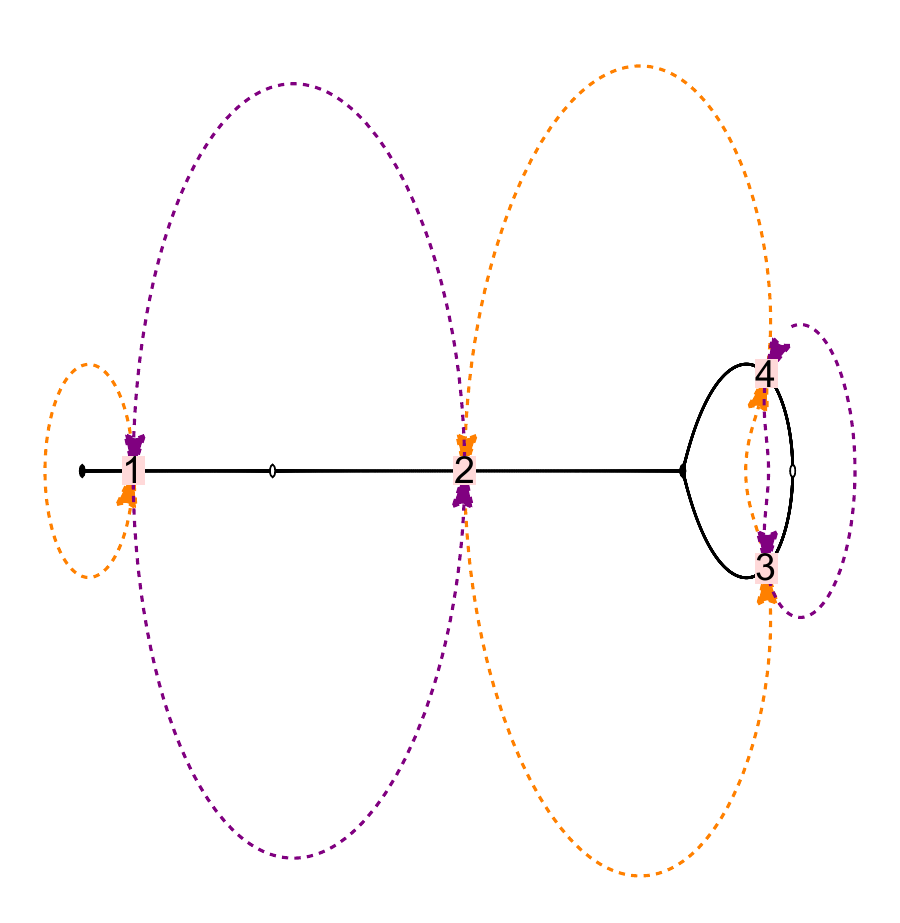}
    \caption{The dessin associated to $\Gamma_0(3)$ with passport $\{1^13^1,2^2,1^13^1\}$. Here, the numbers are the labels of the edges, and the orange (purple) cycles indicate the permutations around black (white) vertices.}\label{monodromyex}
\end{figure}
It is then easy to see that the monodromy group $G$ is generated by $\sigma_0=\{(1),(234)\}$ and $\sigma_1=\{(12),(34)\}$. This is a subgroup of $\mathfrak{S}_{4}$ with $|G|=12$. From $\sigma_\infty\circ\sigma_1\circ\sigma_0=1$, we get $\sigma_\infty=\{(142)\}$.

Now for instance, let us choose a point on edge 2 as reference point. Then the monodromy for the $7$-brane associated to the internal face can be chosen to correspond to the permutation $\sigma^a=(234)$ while the (total) monodromy associated to the leftmost black vertex can be chosen to correspond to $\sigma^b=(12)$. As a result, the (total) monodromy for the $7$-branes associated to the external face is $\sigma^c=(2143)$ so that $\sigma^c\circ\sigma^b\circ\sigma^a=1$. It is obvious that $\sigma^a\in\sigma_0$, $\sigma^b\in\sigma_1$ and $\sigma^c=(\sigma^b\circ\sigma^a)^{-1}$ all belong to $G$.

As the choice for $(p,q)$ is not unique, alternatively we may also choose for example $\sigma^a=(234)(34)=(23)$ and $\sigma^b=(12)$. Then $\sigma^c=(213)$.
\end{example}

\paragraph{A comment on F-theory on elliptically fibred K3} It is well-known that the compactification of F-theory on an elliptically fibred K3 surface is dual to heterotic string theory compactified on $\mathbb{T}^2$. In this setting, the elliptic fibre is still $y^2=x^3+f(k)x+g(k)$ with $k\in\mathbb{P}^1$, but now the degrees of $f$ and $g$ become 8 and 12 respectively. Hence, the number of 7-branes is 24. Although the graph consisting of edges connecting black and white vertices may not be a dessin or even be bipartite any more, the above discussions should still apply following similar methods.

As observed in \cite{Stienstra:2005wy}, certain expansions with respect to the modular Mahler measure coincides with the series for certain Gromov-Witten invariants in the context of F-theory on the toric CY threefolds. We shall leave this to future work.

\section*{Acknowledgments}
We are indebted to the referee, who pointed out the key flaws and kindly brought many more connections to various mathematical objects to our attention, for his/her great patience and deep insight. JB is supported by a CSC scholarship. YHH would like to thank STFC for grant ST/J00037X/1. The research of AZ has been supported by
the French “Investissements d’Avenir” program, project ISITE-BFC (No. ANR-15-IDEX-0003), and EIPHI Graduate School (No. ANR-17-EURE-0002).

\appendix

\section{Maximally and Minimally Tempered Newton Polynomials}\label{Pzw}
In Table \ref{Pzwmaxmin}, we list all the maximally and minimally tempered Newton polynomials for the reflexive polygons.
\begingroup
\renewcommand{\arraystretch}{1.5}
\begin{longtable}{|c|c|c|} \hline
Polygon & Maximally tempered polynomial & Minimally tempered polynomial \\ \hline
No.1 & $k-\frac{w^2}{z}-\frac{z^2}{w}-\frac{3 w}{z}-\frac{3 z}{w}-\frac{1}{w z}-3 w-\frac{3}{w}-3 z-\frac{3}{z}$ & $k-\frac{w^2}{z}-\frac{z^2}{w}-\frac{1}{w z}$ \\ \hline
No.2 & $k-\frac{w^3}{z}-\frac{4 w^2}{z}-\frac{6 w}{z}-\frac{z}{w}-\frac{1}{w z}-2 w-\frac{2}{w}-\frac{4}{z}$ & $k-\frac{w^3}{z}-\frac{z}{w}-\frac{1}{w z}$ \\ \hline
No.3 & $k-\frac{w^2}{z}-\frac{3 w}{z}-\frac{z}{w}-\frac{1}{w z}-2 w-\frac{2}{w}-z-\frac{3}{z}$ & $k-\frac{w^2}{z}-\frac{z}{w}-\frac{1}{w z}-z$ \\ \hline
No.4 & $k-w z-\frac{z}{w}-\frac{w}{z}-\frac{1}{w z}-2 w-\frac{2}{w}-2 z-\frac{2}{z}$ & $k-w z-\frac{z}{w}-\frac{w}{z}-\frac{1}{w z}$ \\ \hline
No.5 & $k-\frac{w^2}{z}-\frac{3 w}{z}-\frac{1}{w z}-2 w-\frac{1}{w}-z-\frac{3}{z}$ & $k-\frac{w^2}{z}-\frac{1}{w z}-\frac{1}{w}-z$ \\ \hline
No.6 & $k-\frac{w^2}{z}-\frac{2 w}{z}-\frac{z}{w}-2 w-\frac{1}{w}-z-\frac{1}{z}$ & $k-\frac{w^2}{z}-\frac{z}{w}-\frac{1}{w}-z-\frac{1}{z}$ \\ \hline
No.7 & $k-\frac{w^2}{z}-\frac{3 w}{z}-\frac{1}{w z}-2 w-z-\frac{3}{z}$ & $k-\frac{w^2}{z}-\frac{1}{w z}-z$\\ \hline
No.8 & $k-\frac{w^2}{z}-\frac{2 w}{z}-2 w-\frac{1}{w}-z-\frac{1}{z}$ & $k-\frac{w^2}{z}-\frac{1}{w}-z-\frac{1}{z}$ \\ \hline
No.9 & $k-\frac{w}{z}-\frac{1}{w z}-w-\frac{1}{w}-z-\frac{2}{z}$ & $k-\frac{w}{z}-\frac{1}{w z}-w-\frac{1}{w}-z$ \\ \hline
No.10 & $k-\frac{z}{w}-\frac{w}{z}-w-\frac{1}{w}-z-\frac{1}{z}$ & $k-\frac{z}{w}-\frac{w}{z}-w-\frac{1}{w}-z-\frac{1}{z}$ \\ \hline
No.11 & $k-\frac{w}{z}-\frac{1}{w z}-\frac{1}{w}-z-\frac{2}{z}$ & $k-\frac{w}{z}-\frac{1}{w z}-\frac{1}{w}-z$\\ \hline
No.12 & $k-\frac{1}{w z}-w-\frac{1}{w}-z-\frac{1}{z}$ & $k-\frac{1}{w z}-w-\frac{1}{w}-z-\frac{1}{z}$ \\ \hline
No.13 & $k-\frac{w}{z}-\frac{1}{w z}-z-\frac{2}{z}$ & $k-\frac{w}{z}-\frac{1}{w z}-z$ \\ \hline
No.14 & $k-\frac{1}{w z}-w-\frac{1}{w}-z$ & $k-\frac{1}{w z}-w-\frac{1}{w}-z$ \\ \hline
No.15 & $k-w-\frac{1}{w}-z-\frac{1}{z}$ & $k-w-\frac{1}{w}-z-\frac{1}{z}$ \\ \hline
No.16 & $k-z-w-\frac{1}{z w}$ & $k-z-w-\frac{1}{z w}$ \\ \hline
\caption{The maximally and minimally tempered Newton polynomials for reflexive polygons.}\label{Pzwmaxmin}
\end{longtable}
\endgroup

\section{Elliptic Curves for Minimally Tempered Coefficients}\label{minimally}
Although not as physically interesting as the maximally tempered coefficients, let us list the results for minimally tempered coefficients for comparison and reference.
As the reflexive polygons No.10, 12, 14, 15 and 16 do not have any boundary points other than vertices, the minimally tempered coefficients coincide with the maximially tempered coefficients. Hence, we will not repeat their results here.

As shown in Table \ref{ellipticminimally}, the elliptic curves for minimally tempered coefficients are not the same for specular duals. The reason is that these coefficients do not encode all the information of the corresponding (numbers of) perfect matchings.
\begingroup
\renewcommand{\arraystretch}{1.5}
\begin{longtable}{c||c|c|c|c} \hline
Polygon & No.1 & No.2 & No.3 & No.4 \\ \hline
$a(k)$ & $-\frac{9}{2}k$ & $-4$ & $-3$ & $\frac{1}{3}k^2-\frac{16}{3}$ \\ \hline
$b(k)$ & $-\frac{5}{8}k^3-\frac{27}{4}$ & $-\frac{2}{3}k^2$ & $-\frac{3}{4}k^2-2$ & $-\frac{1}{36}k^4-\frac{4}{9}k^2+\frac{128}{27}$ \\ \hline
$j(k)$ & $\frac{k^3(k^3+216)^3}{(k^3-27)^3}$ & $\frac{(k^4+192)^3}{(k^4-64)^2}$ & $\frac{(k^4+144)^3}{k^2(k^2-16)(k^2+9)^2}$ & $\frac{(k^4-16k^2+256)^3}{k^4(k^2-16)^2}$ \\ \hline\hline
Polygon & \multicolumn{2}{c|}{No.5} & \multicolumn{2}{c}{No.6} \\ \hline
$a(k)$ & \multicolumn{2}{c|}{$1$} & \multicolumn{2}{c}{$\frac{1}{6}k^2+\frac{1}{2}k+\frac{5}{3}$} \\ \hline
$b(k)$ & \multicolumn{2}{c|}{$-\frac{1}{12}k^2-k-1$} & \multicolumn{2}{c}{$-\frac{1}{72}k^4-\frac{1}{24}k^3-\frac{1}{9}k^2-\frac{5}{6}k-\frac{125}{108}$} \\ \hline
$j(k)$ & \multicolumn{2}{c|}{$\frac{(k^4-48)^3}{k^7+k^6+k^4-72k^3-504k^2-864k-496}$} & \multicolumn{2}{c}{$\frac{(k^4-8k^2-24k-80)^3}{(k^2+4k+5)^2(k^3-6k^2+12k-35)}$} \\ \hline\hline
Polygon & No.7 & No.8 & \multicolumn{2}{c}{No.9} \\ \hline
$a(k)$ & $0$ & $\frac{1}{2}k^2+k$ & \multicolumn{2}{c}{$\frac{1}{6}k^2+k+\frac{2}{3}$} \\ \hline
$b(k)$ & $-1$ & $-\frac{1}{24}k^4-\frac{1}{12}k^3+\frac{1}{4}k^2+k+1$ & \multicolumn{2}{c}{$-\frac{1}{72}k^4-\frac{1}{12}k^3-\frac{1}{36}k^2+\frac{1}{3}k-\frac{7}{27}$} \\ \hline
$j(k)$ & $\frac{k^{12}}{k^6-432}$ & $\frac{(k^4-8k^2-32)^3}{k^6-11k^4-32k^2-256}$ & \multicolumn{2}{c}{$\frac{(k^4-8k^2-48k-32)^3}{(k+1)^2(k^4-8k^2-64k-48)}$} \\ \hline\hline
Polygon & \multicolumn{2}{c|}{No.11} & \multicolumn{2}{c}{No.13} \\ \hline
$a(k)$ & \multicolumn{2}{c|}{$\frac{1}{2}k+1$} & \multicolumn{2}{c}{$1$} \\ \hline
$b(k)$ & \multicolumn{2}{c|}{$-\frac{1}{24}k^3-\frac{1}{12}k^2+\frac{1}{4}$} & \multicolumn{2}{c}{$-\frac{1}{12}k^2$} \\ \hline
$j(k)$ & \multicolumn{2}{c|}{$\frac{(k^4-24k-48)^3}{k^5+k^4+k^3-30k^2-96k-91}$} &  \multicolumn{2}{c}{$\frac{(k^4-48)^3}{k^4-64}$} \\ \hline
\caption{The data of the elliptic curves $y^2=x^3+fx+g$ and $j$-invariants for reflexive polygons (with minimally tempered coefficients). Again, we have $f=-\frac{1}{48}k^4+a(k)$ and $g=\frac{1}{864}k^6+b(k)$ here.}\label{ellipticminimally}
\end{longtable}
\endgroup

\begin{figure}[h]
    \centering
    \begin{subfigure}{4.5cm}
		\centering
		\includegraphics[width=4cm]{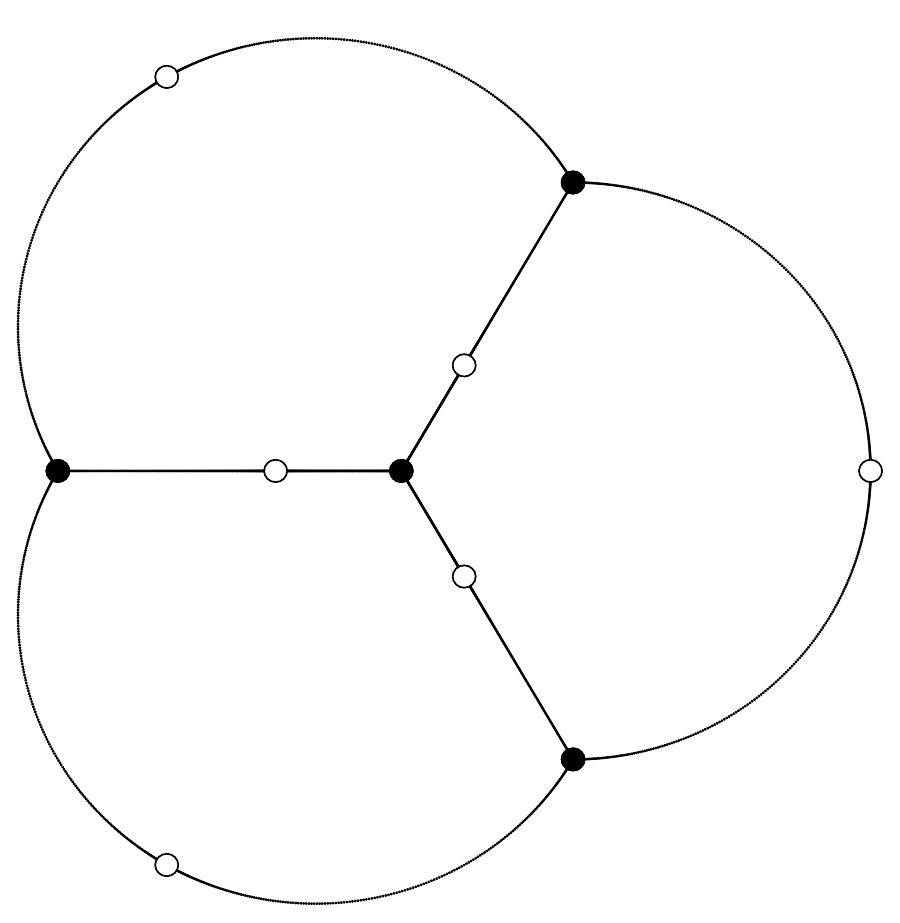}
		\caption{No.1: $\Gamma(3)$}
	\end{subfigure}
	\begin{subfigure}{4.5cm}
		\centering
		\includegraphics[width=4cm]{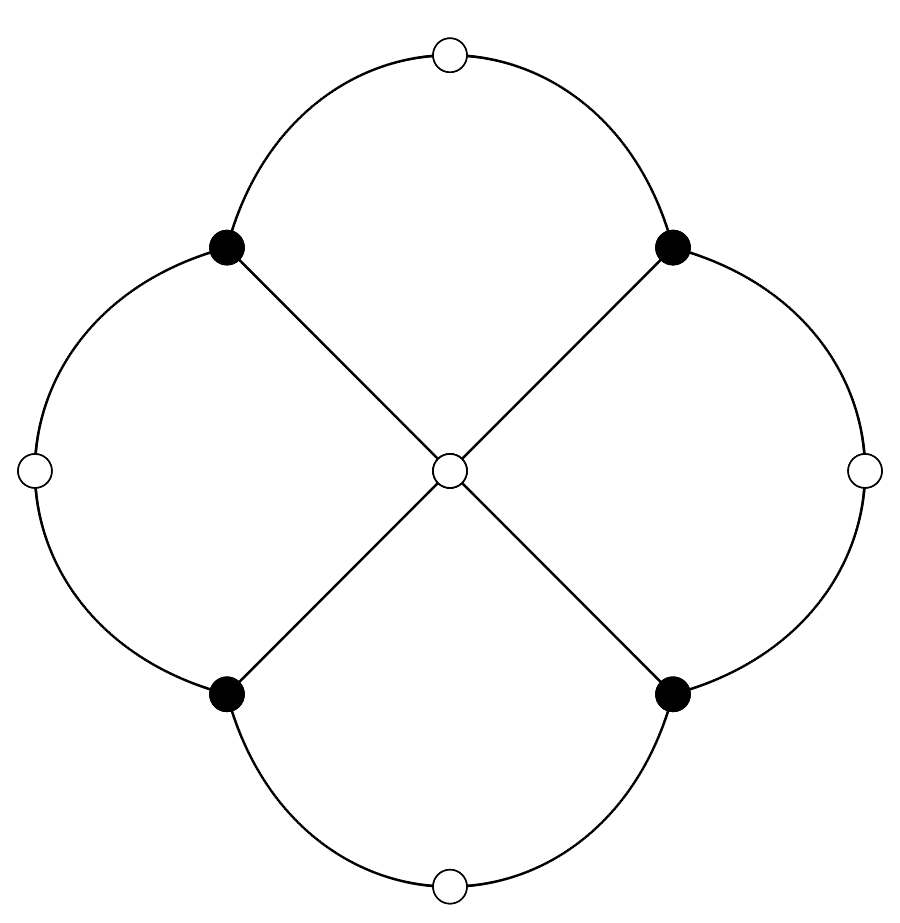}
		\caption{No.2}
	\end{subfigure}
	\begin{subfigure}{4.5cm}
		\centering
		\includegraphics[width=4cm]{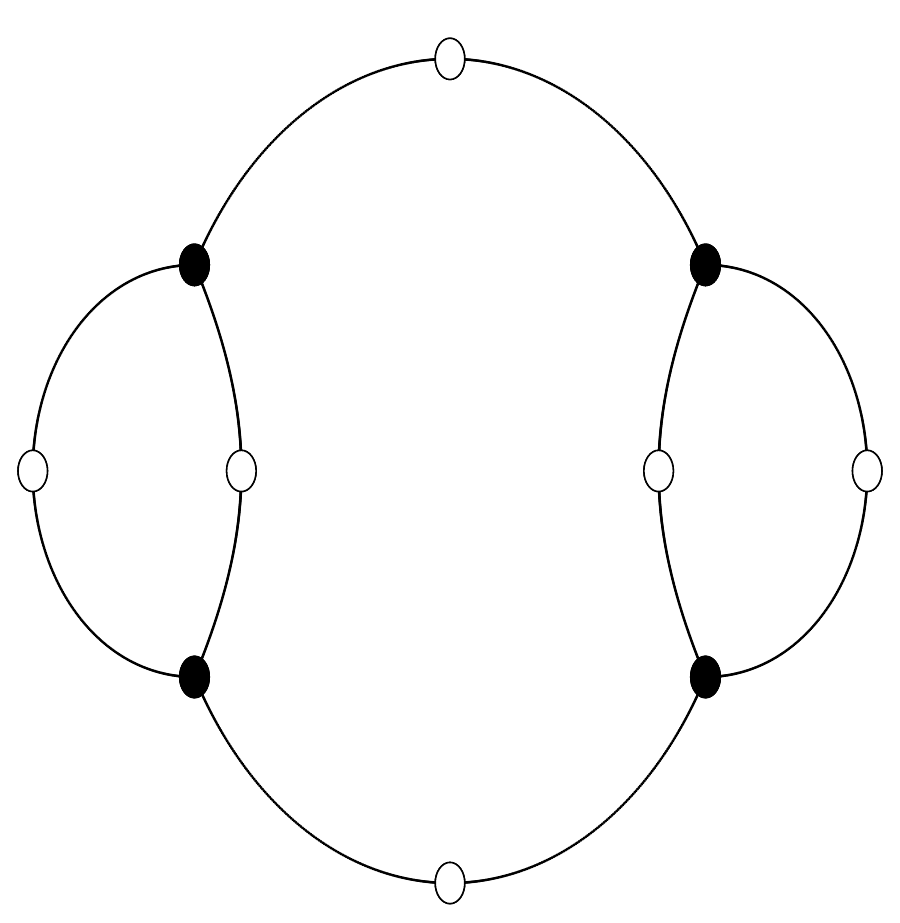}
		\caption{No.4: $\Gamma_0(4)\cap\Gamma(2)$}
	\end{subfigure}
	\begin{subfigure}{4.5cm}
		\centering
		\includegraphics[width=4cm]{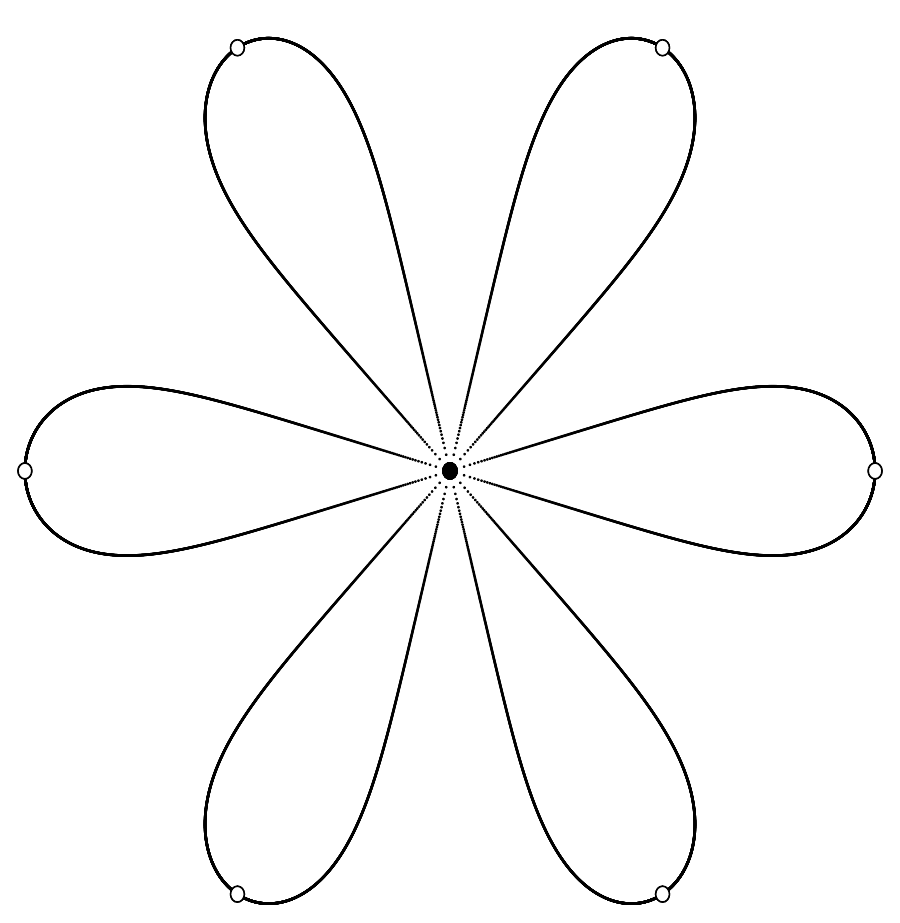}
		\caption{No.7}
	\end{subfigure}
	\begin{subfigure}{4.5cm}
		\centering
		\includegraphics[width=4cm]{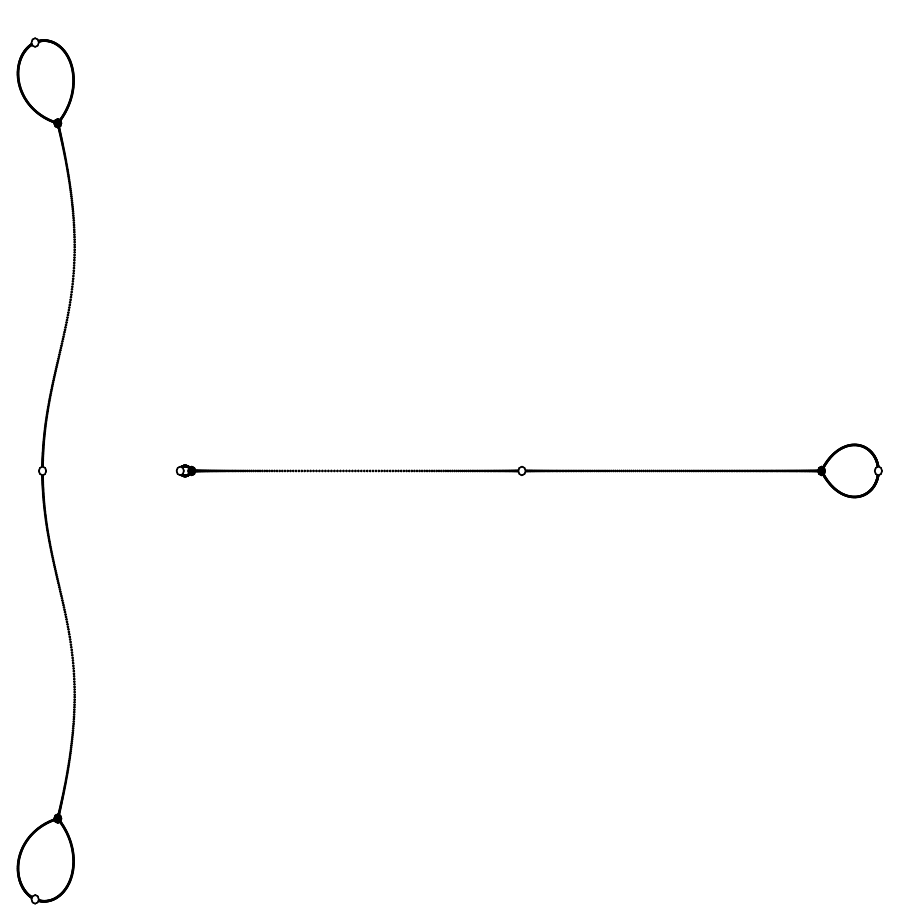}
		\caption{No.9: $\Gamma_0(3)\times\Gamma_0(4)$}
	\end{subfigure}
	\begin{subfigure}{4.5cm}
		\centering
		\includegraphics[width=4cm]{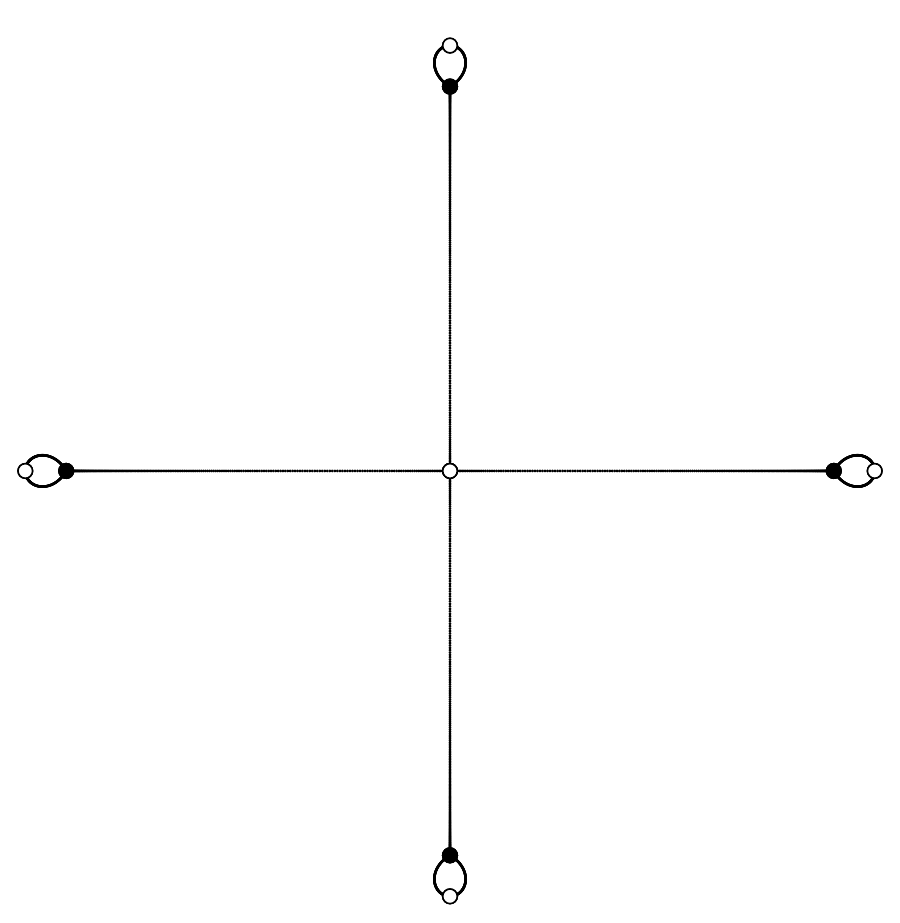}
		\caption{No.13}
	\end{subfigure}
    \caption{The dessins for reflexive polygons with minimally tempered coefficients. As listed, some of them correspond to certain congruence subgroups (as coset graphs).}\label{dessinminimally}
\end{figure}

In Figure \ref{dessinminimally}, we list the plots obtained from $j(k)/1728$. It turns out that No.3, 5, 6, 8, 9, 11 are not Belyi (which are the reflexive polygons in the second and fourth columns in Figure \ref{refpolygons}). Although No.9 is not Belyi, it gives rise a disconnected graph composed of two dessins associated two congruence subgroups. Therefore, we also include it in Figure \ref{dessinminimally}.

One may also compute the Mahler measures for the Newton polynomials $P(z,w)=k-p(z,w)$ with those minimally tempered coefficients as series of $k$. We will not list them here, but we would like to point out two properties:
\begin{itemize}
    \item There are several (but not all) reflexive pairs giving the same Mahler measures. These pairs are No.1\&16, No.2\&13, No.4\&15 (plus the self-dual ones). The reason is that the vertices of the polygons in each pair are related by some GL$(2,\mathbb{Z})$ transformation (while the other reflexive duals are not). This can be seen by quotient gradings on the lattice or direct computations of Pl\"ucker coordinates. As the minimally tempered coefficients only contain the vertices, this then follows from the fact that Mahler measure is GL$(2,\mathbb{Z})$ invariant.
    \item There are four classes of polygons whose Mahler measures can be expressed compactly using some generalized hypergeometric functions ${}_4F_3$. Likewise, their $u_0$ are also simply some hypergeometric functions ${}_2F_1$. These four classes are classified in \cite{villegas1999modular}. It turns out that the four classes are precisely No.1\&16, No.2\&13, No.4\&15 and the self-dual No.7.
\end{itemize}

Although not all dessins  are associated to congruence subgroups, we may still compute the modular expansions for the $k$ parameters and check if they give rise to any Hauptmoduln. Here we give three examples of different types. The detailed steps can be found in \cite{Stienstra:2005wy,samart2014mahler}.

\paragraph{Example 1: No.1} As this is the same as the case for dP$_0$ (No.16), we have computed that $k^3=27+\left(\frac{\eta(\tau)}{\eta(3\tau)}\right)^{12}$. This is a Hauptmodul for $\Gamma_0(3)$. In particular, the congruence subgroup associated to the dessin in this case is $\Gamma(3)$, which is a subgroup of $\Gamma_0(3)$.

\paragraph{Example 2: No.13} We have
\begin{equation}
    k^4=q^{-1}+40+276q-2048q^2+11202q^3+\dots=64+\left(\frac{\eta(\tau)}{\eta(2\tau)}\right)^{24},
\end{equation}
where the second equality is checked perturbatively. This is a Hauptmodul for $\Gamma_0(2)$. On the other hand, the crossing dessin does not correspond to any congruence subgroup. By removing the white vertices (or black vertices), this does not even seem to be a coset graph for any group either.

\paragraph{Example 3: No.7} We have
\begin{equation}
    k^6=864\left(1-\frac{E_6(\tau)}{E_4^{3/2}(\tau)}\right),
\end{equation}
where $E_4(\tau)=1+240\sum\limits_{n=1}^\infty\frac{n^3q^n}{1-q^n}$ and $E_6(\tau)=1-540\sum\limits_{n=1}^\infty\frac{n^5q^n}{1-q^n}$ are the Eisenstein series. This is not known to be a Hauptmodul of any genus-0 congruence subgroup. On the other hand, the flower dessin does not correspond to any congruence subgroup either. By removing the white vertices, however, it could be viewed as a coset graph associated to any group with 6 generators (and the subgroup being itself). Incidentally, there are two things worth noting:
\begin{itemize}
    \item As given in \cite{FrickeKlein}, we have
    \begin{equation}
        E_4^{1/4}(\tau)={}_2F_1\left(\frac{1}{12},\frac{5}{12};1;\frac{1728}{j}\right)
    \end{equation}
    in terms of $j$-invariant (cf. \S\ref{hauptmoduln}).
    \item The $q$-series expansion for $E_4(\tau)$ has $a_n=240$ for $n\geq1$. It turns out that $2a_n=480$ are the GW invariants in the first row of Table 1 given in \cite{Klemm:1996hh} (cf. \S\ref{GW}).
\end{itemize}

\paragraph{The del Pezzo surfaces of degrees 1 and 2} The quantum periods for del Pezzo surfaces of degrees 1 and 2 are also provided in \cite{coates2016quantum}:
\begin{equation}
    G_{X_1}(t)=\text{e}^{-60t}\sum_{n=0}^{\infty}\frac{(6n)!}{n!(2n)!(3n)!}\frac{t^n}{n!},\quad G_{X_2}(t)=\text{e}^{-12t}\sum_{n=0}^{\infty}\frac{(4n)!}{(n!)^2(2n)!}\frac{t^n}{n!}.
\end{equation}
They can be written in terms of the following integrals:
\begin{align}
    &G_{X_1}(t)=\frac{1}{(2\pi i)^2}\int_{|z|=|w|=1}\exp\left(\left(-60+p_1(z,w)^6\right)t\right)\frac{\text{d}z}{z}\frac{\text{d}w}{w},\label{GX1}\\
    &G_{X_2}(t)=\frac{1}{(2\pi i)^2}\int_{|z|=|w|=1}\exp\left(\left(-12+p_2(z,w)^4\right)t\right)\frac{\text{d}z}{z}\frac{\text{d}w}{w},\label{GX2}
\end{align}
where
\begin{equation}
    p_1(z,w)=z^{-1}w^{-1}+z+z^{-1}w^2,\quad p_2(z,w)=z^{-1}w^{-1}+z+z^{-1}w.
\end{equation}
They are exactly the expressions for the reflexive polygons 7 and 13 respectively with minimally tempered coefficients.

\paragraph{The Beauville pencils I and II} Now, let us check that the Beauville pencils I and II are also related to reflexive polygons, but with minimally tempered coefficients. For I, it is straightforward to see that this is exactly the reflexive polygon No.1 with minimally tempered coefficients. The congruence subgroups for the Beauville pencil and the dessin are both $\Gamma(3)$. Now for II, this gives the Newton polynomial for No.13 (but with neither maximally nor minimally tempered coefficients). However, we find that the Weierstrass form is the same as the reflexive polygon No.4 with minimally tempered polynomial. In other words, they parametrize the same elliptic curve. Therefore, this belongs to the same group as No.4 with minimally tempered coefficients does. Recall that the congruence subgroup associated to this Beauville pencil is $\Gamma_1(4)\cap\Gamma(2)$ as listed in Table \ref{Beauville}. This is indeed a subgroup of $\Gamma_0(4)\cap\Gamma(2)$ for the dessin as shown in Figure \ref{dessinminimally}. This completes the verification of Propostition \ref{Beauvilleprop}.

\paragraph{Singular fibres} For these minimally tempered coefficients, the singular fibre types can also be obtained from the information of $f,g$ and $\Delta$. Let us list them in Table \ref{singfibsmin}. For No.7 and No.13, they correspond to the del Pezzo surfaces of degrees 1 and 2 respectively. Now, the degree does not coincide with the fibre type $I_N$. Instead,
\begin{equation}
    N=r\cdot\text{degree},
\end{equation}
where $r$ is the power of $p(z,w)$ (namely, $p(z,w)^r$) appeared in the integrals \eqref{GX1} and \eqref{GX2}.
\begingroup
\renewcommand{\arraystretch}{1.5}
\begin{longtable}{|c|c|} \hline
 Reflexive polygon & Singular fibres \\ \hline
No.1 & $I_3,I_3,I_3$ \\ \hline
No.2 & $I_4,I_2,I_2,I_2,I_2$ \\ \hline
No.3 & $I_4,I_2,I_2,I_2,I_1,I_1$ \\ \hline
No.4 & $I_4,I_4,I_2,I_2$\\ \hline
No.5 & $I_5,I_1,I_1,I_1,I_1,I_1,I_1,I_1$ \\ \hline
No.6 & $I_5,I_2,I_2,I_1,I_1,I_1$ \\ \hline
No.7 & $I_6,I_1,I_1,I_1,I_1,I_1,I_1$ \\ \hline
No.8 & $I_6,I_3,I_2,I_1$ \\ \hline
No.9 & $I_6,I_1,I_1,I_1,I_1,I_1,I_1$ \\ \hline
No.11 & $I_7,I_1,I_1,I_1,I_1,I_1$ \\ \hline
No.13 & $I_8,I_1,I_1,I_1,I_1$ \\ \hline
\caption{The singular fibre types determined from $f,g$ and $\Delta$ associated to the reflexive polygons with minimally tempered coefficients. We do not repeat the cases where the maximally and minimally tempered coefficients coincide. The fibre at $k=\infty$ is listed first in each row.}\label{singfibsmin}
\end{longtable}
\endgroup

\section{Mahler Measure and $j$-Invariant}\label{mj}
As the Mahler measures and dessins are connected to each other, it should be possible to write $m(P)$ in terms of $j$. Let us first start with a rather general definition of periods introduced in \cite{kontsevich2001periods}:
\begin{definition}
A \textbf{period} is a complex number whose real and imaginary parts are values of absolutely convergent integrals of rational functions with rational coefficients, over domains in $\mathbb{R}$ given by polynomial inequalities with rational coefficients.
\end{definition}
As a matter of fact, the set of periods, which is countable, form an algebra under the usual sum and product operations.
Famous constants such as $\pi$ can be shown to be periods.
In particular, when the Newton polynomial has rational coefficients, {\it the Mahler measure is a period} \cite{kontsevich2001periods}. For those considered in this paper, i.e., $P(z,w)=k-p(z,w)$ with \emph{any} tempered coefficients, this means $m(P)$ is a period when $k\in\mathbb{Q}$.

An important theorem in \cite{kontsevich2001periods} says that
\begin{theorem}
Consider $\textup{SL}(2,\mathbb{Z})$ or any of its subgroup of finite index. Let $f(z)$ be a modular form (either holomorphic or meromorphic) of some positive weight $\mathfrak{w}$ and let $t(z)$ be a modular function under the action of the group. Then $F(t(z))\equiv f(z)$, which is multi-valued, satisfies a homogeneous linear differential equation of order $(\mathfrak{w}+1)$, $\sum\limits_{n=0}^{\mathfrak{w}+1}a_nF^{(n)}(t(z))=0$ with $a_n$ algebraic functions of $t(z)$.\label{diffeq}
\end{theorem}

Since $\lambda$, $u_0$ and $e$ are modular forms of weights 0, 1 and 3 respectively (though with singularities), and since $j(\tau)$ is a modular function, we have
\begin{corollary}
The modular forms $\lambda(j(\tau))$, $u_0(j(\tau))$ and $e(j(\tau))$ satisfy linear differential equations (with respect to $j$) of order $1$, $2$ and $4$ respectively.
\end{corollary}
Recall that $e$ generates the coefficients of $m(P)$ in $q$-series. It would therefore be reasonable to expect certain relations between Mahler measures and $j$-invariants.

Another crucial result in \cite{kontsevich2001periods} says that
\begin{theorem}
Let $f(z)$ be a modular form of positive weight $\mathfrak{w}$ and let $t(z)$ be a modular function, both defined over $\overline{\mathbb{Q}}$. Then $\forall z_0\in\mathfrak{h}$ for which $t(z_0)$ is algebraic, $\pi^{\mathfrak{w}}f(z_0)$ is a period.
\end{theorem}
We may now apply this theorem to the modular forms in our paper.
\begin{corollary}
When $j(\tau)$ is algebraic, $\lambda(j(\tau))$, $\pi u_0(j(\tau))$ and $\pi^3e(j(\tau))$ are periods.
\end{corollary}
Moreover, when $j\in\overline{\mathbb{Q}}$, we also learn that $q$ is transcendental following \cite{Diaz2001mahlerconj}. Since $m(P)$ is a period when $P\in\mathbb{Q}[z^{\pm1},w^{\pm1}]$ and $j$ is a rational function of $k$, we learn that $m(P)$ is a period if $j$ is rational. In fact, we can extend this to $j$ being algebraic. This is because $m(P)$ is a sum over $\lambda^n/n$ with integer coefficients\footnote{This can be seen from $u_0$ as its coefficients are integers that count F-term relations.}. Now this follows from $\lambda$ being a period and that periods form an algebra with countably many elements\footnote{The infinite sum should be well-defined in the algebra assuming that the Mahler measure converges.}. Hence, we conclude that
\begin{proposition}
The Mahler measure $m(P)$ is a period if $j$ is algebraic.\label{mjperiodalgebraic}
\end{proposition}

Since $\lambda$ and $u_0$ satisfy certain differential equations, that is,
\begin{equation}
    \lambda'=\alpha_0\lambda,\quad u_0''+\alpha_1u_0'+\alpha_2u_0=0,
\end{equation}
where $f'$ denotes the derivative with respect to $j$ and $\alpha_{0,1,2}$ are differentiable algebraic functions of $j$, we can use the Mahler flow equation
\begin{equation}
    \frac{\text{d}m}{\text{d}\log\lambda}=\lambda\frac{\text{d}m}{\text{d}\lambda}=-u_0
\end{equation}
to get
\begin{equation}
    \lambda\frac{\text{d}m}{\text{d}j}\frac{1}{\lambda'}=\frac{1}{\alpha_0}m'=-u_0.
\end{equation}
Plugging this into the Picard-Fuchs equation for $u_0$ (with respect to $j$) yields
\begin{equation}
    m'''(j)+\left(\alpha_1-\frac{2\alpha_0'}{\alpha_0}\right)m''(j)+\left(\alpha_2-\frac{\alpha_1\alpha_0'}{\alpha_0}+\frac{2\left(\alpha_0'\right)^2}{\alpha_0^2}-\frac{\alpha_0''}{\alpha_0}\right)m'(j)=0.\label{mjDE}
\end{equation}

\begin{figure}[h]
    \centering
    \includegraphics[width=5cm]{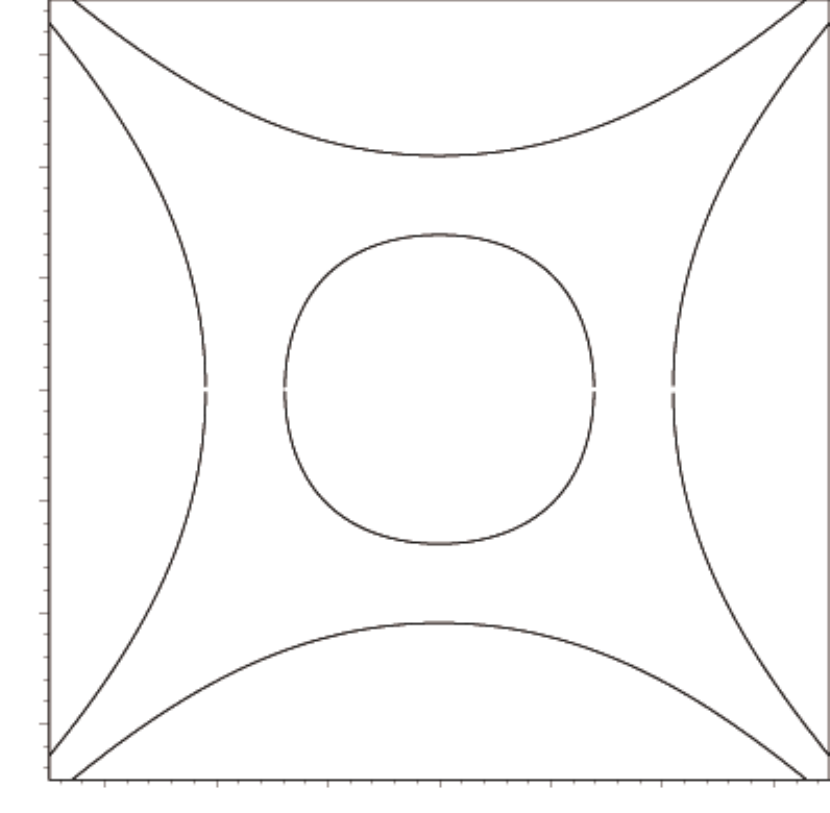}
    \includegraphics[width=4.5cm]{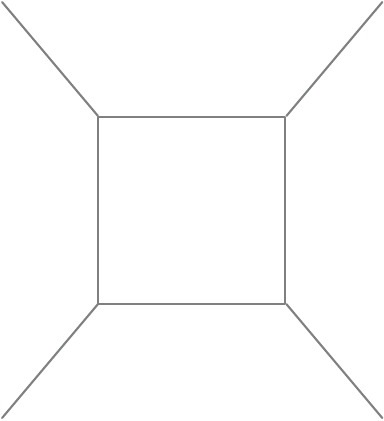}
    \caption{Left: The amoeba of the polygon No.15 with $p(z,w)=k-z-w-z^{-1}-w^{-1}$ for finite $k>4$ (for $k\leq4$, there is no hole in the middle). Right: The amoeba of the same polygon for $k\rightarrow\infty$, which we shall refer to as the tropical limit coming from tropical geometry.}\label{amoeba}
\end{figure}

\paragraph{Tropical limit} For convenience, let us call the limit $k\rightarrow\infty$ the tropical limit due to the fact that the amoeba\footnote{The amoeba is the set $\mathcal{A}(P)=\{(\log|z|,\log|w|)|P(z,w)=0\}$ on the real plane.} of the Newton polynomial tends to its tropical spine, namely the dual of the toric diagram, when $k\rightarrow\infty$. An illustration can be found in Fig.~\ref{amoeba}.
Recall that $m\sim\log k$ in the tropical limit. Likewise, we have $j\rightarrow\infty$ as $j(k)$ is a rational function $\frac{f_1(k)}{f_2(k)}$ with $\deg(f_1)>\deg(f_2)$. More precisely, $j\sim k^n$ in the tropical limit, where $n=\deg(f_1)-\deg(f_2)$ is the power for the external face in the passport of the corresponding dessin. Hence, in the tropical limit,
\begin{equation}
    m\sim\frac{1}{n}\log j.\label{mjtrop}
\end{equation}
One may wonder whether for any $j$, $m(j)$ can be expressed by further adding a sum of $-c_l/j^l$ where $l\geq2$ are integers and $c_l$ are some coefficients. However, due to the multi-valuedness of $k$ as a function of $j$, this would only be valid on one branch. Indeed, $j$ could still diverge for some finite $k$ while $m(j)$ would remain finite in this case.

\begin{remark}
Using Mahler flow equation and Picard-Fuchs equation \eqref{PF}, we may also write $m(P)$ as a differential equation with respect to $\lambda$ as $\lambda$ is of weight $0$ (under the monodromy of Picard-Fuchs equation). Then the differential equation reads
\begin{equation}
    \lambda A\frac{\textup{d}^3m(\lambda)}{\textup{d}\lambda^3}+(\lambda B+2A)\frac{\textup{d}^2m(\lambda)}{\textup{d}\lambda^2}+(\lambda C+B)\frac{\textup{d}m(\lambda)}{\textup{d}\lambda}=0.
\end{equation}
\end{remark}

To conclude this subsection, we should emphasize that our discussions on the relations between between $m(P)$ and $j$ are still very shallow here. For instance, the differential equation \eqref{mjDE} is more like a proof of concept. It would only be useful when one knows how to determine the coefficients in the equation. Moreover, the connection of the Mahler measure to the periods is a very big topic, and it should involve various objects and concepts in mathematics such as $L$-functions, mixed motives \cite{deninger1997deligne} etc. We therefore leave deeper investigations for future works.

\section{Mahler Flow and the $\tau_{R,G,B}$ Conjecture}\label{taubtaur}
There has been a long puzzle about the nature of brane tilings as bipartite graphs on $\mathbb{T}^2$ \cite{Jejjala:2010vb,Hanany:2011ra,Hanany:2011bs,He:2012xw,Bao:2021ohf}.
On the one hand, they could be interpreted as dessins\footnote{They are embedded on $\mathbb{T}^2$ instead of $\mathbb{P}^1$ compared to the dessins discussed in this paper so far.} on $\mathbb{T}^2$, acquiring a complex structure called $\tau_B$ (the subscript $B$ indicates its origin from Belyi) which is that of $\mathbb{T}^2$ as an elliptic curve.

On the other hand, the R-charges in the quiver theory obtained from the isoradial brane tiling correspond to angles of the faces in the tiling\footnote{In fact, we may also treat non-isoradial tilings as ``isoradial'' tilings in a similar manner if we allow zero or negative angles. See \cite{Bao:2021fjd} for more details.}. The R-charges would then determine the complex structure $\tau_R$ on the torus which supports the tiling.
It would be natural to suspect that the two complex structures would coincide as conjectured in \cite{Jejjala:2010vb}. However, as later discussed in \cite{Hanany:2011ra,Hanany:2011bs}, counterexamples exist and this $\tau_B=\tau_R$ conjecture does not hold in general.

Furthermore, there is a third complex parameter called $\tau_G$ coming directly from the geometry of the CY3 singularity corresponding to the toric diagram. In particular, a U$(1)^2$ subgroup of the $\mathbb{T}^3$-action would leave the K\"ahler form and holomorphic 3-form invariant. When the CY space is viewed as a special Lagrangian fibration, the U$(1)^2$ would then define an invariant part of such fibration, which turns out to be a torus. The metric on this torus, which is the pullback of the metric on the CY singularity, leads to the complex structure $\tau_G$. As studied in \cite{He:2012xw}, $\tau_{R,G,B}$ may sometimes be coincident with each other, but they do not always equal in general.
Notice that when we say $\tau_{R,G,B}$ coincide, this is always up to SL$(2,\mathbb{Z})$ transformation. In practice, we would always compare $j(\tau_{R,G,B})$ as it is modular invariant. Following the study in the above references, we then have the following questions: \emph{When are $j(\tau_{R,G,B})$ equal to each other? What are the relations among them?}

Since $j(k)=\frac{f_1(k)}{f_2(k)}$ where $f_{1,2}(k)$ are polynomials of $k$, the range of $j(k)$ is the whole $\mathbb{C}\sqcup\{\infty\}$. Therefore, no matter what value $j(\tau_{R,G,B})$ takes, there must be at least one $k$ on $\mathbb{P}^1$ such that $j(k)=j(\tau_{R,G,B})$. In the language of \cite{Bao:2021fjd}, we say that they are different points along the Mahler flow. This is true in general, and not just restricted to the reflexive cases. Since we still have the freedom to choose the coefficients for the Newton polynomial even if the polygon is fixed, one may wonder which Mahler flow would be the appropriate choice. As $\tau_R$ originates from R-charges and R-charges are associated to angles in the isoradial (or even non-isoradial) tilings, instead of tempered coefficients, we shall always use the coefficients from the canonical edge weights on the tilings\footnote{Therefore, in $j(k)=\frac{f_1(k)}{f_2(k)}$, $f_{1,2}$ have algebraic coefficients.}. Here, we shall make a conjecture involving only $\tau_R$ and $\tau_B$ based on the observation of all the computed cases. Let us first look at some examples.

\begin{example}
For (chiral) orbifolds of $\mathbb{C}^3$ and of the conifold ($\mathcal{C}$), the three complex structures coincide \cite{Jejjala:2010vb,Hanany:2011ra,Hanany:2011bs,He:2012xw}, due to the hexagonal and square symmetries of the tiling.
For instance, $j(k)=\frac{k^3(k^3-24)^3}{k^3-27}$ for $\textup{dP}_0$, which is a $\mathbb{Z}/3$-orbifold of $\mathbb{C}^3$ and $j(\tau_{R,G,B})=0$. Solving $j(k)=j(k_{R,G,B})$, we find that the $\tau_{R,G,B}$ complex structure is located at
\begin{equation}
    k=0,~2\sqrt[3]{3},~2\sqrt[3]{3}\text{e}^{\pm2\pi i/3}
\end{equation}
on the sphere.

As another example, $j(k)=\frac{(k^4-16k^2+16)^3}{k^2(k^2-16)}$ for $\mathbb{F}_0$, a $\mathbb{Z}/2$-orbifold of $\mathcal{C}$ and $j(\tau_{R,G,B})=1728$. Solving $j(k)=j(k_{R,G,B})$, we find that the $\tau_{R,G,B}$ complex structure is located at
\begin{equation}
    k=\pm2\sqrt{2},~\pm\sqrt{2(4+3\sqrt{2})},~\pm i\sqrt{2(-4+3\sqrt{2})}
\end{equation}
on the sphere.
\end{example}

\begin{example}
Unlike the above example, the suspended pinch point (SPP) and its orbifolds have different $\tau_{R,G,B}$. For instance, $\textup{SPP}/\mathbb{Z}_2$ with action $(0,1,1,1)$ (No.8 in the list of reflexive polygons) has Newton polynomial
\begin{equation}
    P(z,w)=-2A^2zw^2-A^2w^3-A^2z^2w-2ABw^2-B^2z-B^2w+kzw
\end{equation}
for canonical edge weights, where
\begin{equation}
    A=\sin^2\left(\frac{\pi}{2\sqrt{3}}\right),\quad B=\sin\left(\left(1-\frac{1}{\sqrt{3}}\right)\pi\right)\sin\left(\frac{1}{2}\left(1-\frac{1}{\sqrt{3}}\right)\pi\right).
\end{equation}
Therefore,
\begin{equation}
    j(k)=\frac{256C^3k^3(-6-6Ck+C^3k^3)^3}{(-3+Ck)(1+Ck)^3(3+2Ck)^2},
\end{equation}
where
\begin{equation}
    C=\csc\left(\frac{\pi}{2\sqrt{3}}\right)\csc^2\left(\frac{\pi}{\sqrt{3}}\right).
\end{equation}
As computed in \cite{Hanany:2011ra},
\begin{equation}
    j(\tau_B)=\frac{132304644}{5},\quad j(\tau_R)=287496.
\end{equation}
Therefore, the $\tau_B$ complex structure is located at approximately
\begin{equation}
    \begin{split}
        k=&-1.112,~6.909,~-2.898\pm5.439i,~-6.157,~-1.114,~-0.752,~2.2260,\\
        &-0.737\pm0.009i,~3.635\pm5.430i
    \end{split}
\end{equation}
while the $\tau_R$ complex structure is located at approximately
\begin{equation}
    \begin{split}
        k=&-1.107,~3.677,~-1.285\pm2.392i,~-2.892,~-1.112,~-0.785,~2.2261,\\
        &-0.721\pm0.041,~2.006\pm2.351i
    \end{split}
\end{equation}
on the Mahler flow sphere.
\end{example}

Based on the known examples, it seems that the toric diagrams (e.g. $\mathbb{C}$ and $\mathcal{C}$) which satisfy the $\tau_B=\tau_R$ conjecture look more ``symmetric'' than those (e.g. SPP) do not satisfy the $\tau_B=\tau_R$ conjecture. It turns out that the coefficients from canonical weights for those more ``symmetric'' polygons coincide with the maximally tempered coefficients while those from canonical weights for the less ``symmetric'' ones do not agree with the maximally tempered coefficients. Based on the above examples, it is natural to conjecture that
\begin{conjecture}
Up to $\textup{SL}(2,\mathbb{Z})$, the $\tau_B=\tau_R$ condition holds if and only if the maximally tempered coefficients of the Newton polynomial coincide with the coefficients from canonical edge weights on the tiling.\label{taubtaurcondition}
\end{conjecture}

\begin{remark}
Our observation also agrees with the fact that those less ``symmetric'' cases have a more non-trivial $a$-maximization. In particular, it was shown in \cite{Bao:2021fjd} that $a$-maximization is equivalent to maximization of Mahler measure with canonical weights. Therefore, the non-triviality of $a$-maximization can be interpreted as the discrepancy between maximally tempered and canonically weighted coefficients in terms of Mahler measure. 
\end{remark}

For the reflexive polygons in Table \ref{refpolygons}, No.1, 2, 4, 7, 10, 13, 15 and 16 have coincident canonical weights and maximally tempered ones while the two choices are different for the remaining cases. One may check that this satisfy our above discussions.

\section{Mahler Measure and Gromov-Witten Invariants}\label{GW}
When the F-theory is compactified on the toric CY cone associated to a (reflexive) Newton polygon, its effective theory is a closed subsector of the type II compactification. The BPS states of the F-theory compactification should then give a subsector of those in the full type II theory. In \cite{Klemm:1996hh,Lerche:1996ni}, such instanton expansions were computed. In particular, the GW invariants of local vanishing del Pezzo surfaces (independently of the global embedding in the CY spaces where F-theory compactifies) were observed to coincide with certain modular expansions of Mahler measures from the same toric diagrams later in \cite{Stienstra:2005wy}. In a related work \cite{doran2011algebraic}, Mahler measure is discussed in connection with the asymptotic growth rate of the Gromov-Witten invariants in Calabi–Yau hypersurfaces in toric Fano $n$-folds. With certain ansatz, we shall see that the GW invariants of any vanishing 4-cycles could be recovered from such modular expansions from the corresponding toric diagrams according to the dictionary of the two sides.

As an elliptic curve is topologically $\mathbb{T}^2$, the periods are given by $\Tilde{\phi}$ and $\Tilde{\phi}_D$ following the notations of \cite{Lerche:1996ni}. More concretely, given a 6d theory obtained from the F-theory on the CY threefold, we may further compactify it on a $\mathbb{T}^2$ to obtain a 4d supersymmetric gauge theory. Then $\widetilde{\phi}$ is given by $i\phi R_5R_6$, where $\phi$ is the tension of the non-critical string and $R_{5,6}$ are the radii of the two circles in the compactification. The dual period is then $\widetilde{\phi}_D=\partial_{\widetilde{\phi}}\mathcal{F}$ with $\mathcal{F}$ being the prepotential. Then we shall identify the gauge coupling $\partial\Tilde{\phi}_D/\partial\Tilde{\phi}$ with $\tau$ on the modular Mahler measure side, that is,
\begin{equation}
    \tau\sim\frac{\partial\Tilde{\phi}_D}{\partial\Tilde{\phi}}.
\end{equation}

The instanton expansions in \cite{Lerche:1996ni} are worked out at the large complex structure point $c=0$, where $c=\text{e}^{2\pi i\Tilde{\phi}}+\dots$ provides a coordinate on the moduli space. This corresponds to the tropical limit $k\rightarrow\infty$, or equivalently, $q\rightarrow0$. A natural ansatz for the correspondence would then be
\begin{equation}
    c\sim q^{\nu}
\end{equation}
for some $\nu\in\mathbb{Z}^+$.

In order to have the correspondence consistent, our goal is to show that this leads to the correspondence between the Yukawa coupling $C_{\Tilde{\phi}\Tilde{\phi}\Tilde{\phi}}$ in \cite{Lerche:1996ni} and $\frac{\text{d}\log q}{\text{d}m}$ in \cite{Stienstra:2005wy}. In particular, they have the expansions
\begin{equation}
    C_{\Tilde{\phi}\Tilde{\phi}\Tilde{\phi}}=c_0+\sum_{n=1}^\infty\frac{\mathfrak{a}_nn^3q_{\Tilde{\phi}}^n}{1-q_{\Tilde{\phi}}^n},\quad \frac{\text{d}\log q}{\text{d}m}=-1+\sum_{n=1}^\infty\frac{a_nn^3(\text{e}^{-\nu m})^{n}}{1-(\text{e}^{-\nu m})^{n}},
\end{equation}
where $q_{\Tilde{\phi}}:=\text{e}^{2\pi i\Tilde{\phi}}$ and $\nu$, $c_0$ are some positive constants depending on different cases. Therefore, we would like to see that $a_n$ coincides with the GW invariants $\mathfrak{a}_n$ up to the constant $c_0$, that is, $\mathfrak{a}_n=-c_0a_n$.

For these two expansions to match, we need
\begin{equation}
    \text{e}^{-\nu m}\sim q_{\Tilde{\phi}}=\text{e}^{2\pi i\Tilde{\phi}}.
\end{equation}
Indeed, the expansion for $q^{\nu}$ is $q^{\nu}=\text{e}^{-\nu m}+\dots$, which agrees with $c=\text{e}^{2\pi i\Tilde{\phi}}+\dots$. Now, since $m=-2\pi i\tau-\dots$, we have $m\sim-2\pi i\tau$. This would yield $\Tilde{\phi}\sim\nu\tau$. As $u_0\sim1$, we shall further tune the constant factor to be
\begin{equation}
    \Tilde{\phi}\sim-c_0\tau u_0.
\end{equation}
The reason is that with
\begin{equation}
    \Tilde{\phi}_D\sim-\frac{1}{4}\pi ic_0\tau u_1,
\end{equation}
using $u_1\sim u_0\log\lambda\sim u_0\log q=2\pi i\tau u_0$, we can recover
\begin{equation}
    \frac{\partial\Tilde{\phi}_D}{\partial\Tilde{\phi}}=\frac{\partial\Tilde{\phi}_D/\partial\tau}{\partial\Tilde{\phi}/\partial\tau}\sim\frac{1}{4\pi i}\frac{c_0\partial(\tau u_1)/\partial\tau}{c_0u_0}=\tau.
\end{equation}

Now we are ready to show that
\begin{equation}
    C_{\Tilde{\phi}\Tilde{\phi}\Tilde{\phi}}\sim -c_0\frac{\text{d}\log q}{\text{d}m},
\end{equation}
where
\begin{equation}
    C_{\Tilde{\phi}\Tilde{\phi}\Tilde{\phi}}=\frac{\partial^2\Tilde{\phi}_D}{\partial\Tilde{\phi}^2}\sim\frac{\partial\tau}{\partial\Tilde{\phi}}.
\end{equation}
This can be seen as follows. Since $\lambda=q+\dots$, we have
\begin{equation}
    -\frac{\text{d}m}{\text{d}\log q}=-\frac{\text{d}m}{\text{d}\log\lambda}=u_0.
\end{equation}
On the other hand,
\begin{equation}
    \frac{\partial\Tilde{\phi}}{\partial\tau}\sim\frac{\partial(-c_0\tau u_0)}{\partial\tau}=-c_0u_0.
\end{equation}
Thus,
\begin{equation}
    C_{\Tilde{\phi}\Tilde{\phi}\Tilde{\phi}}\sim-\frac{1}{c_0u_0}\sim-c_0\frac{\text{d}\log q}{\text{d}m}.
\end{equation}
Since we are working at the large complex structure point/tropical limit, ``$\sim$'' can be turned into ``$=$''. To summarize, the correspondence of quantities between Mahler measure and GW invariants is listed in Table \ref{dictionary}.
\begin{table}[h]
\renewcommand{\arraystretch}{1.5}
\centering
\begin{tabular}{|c||c|c|c|c|c|c|}
\hline
Mahler & $-c_0\tau u_0$ & $-\frac{1}{4\pi i}c_0\tau u_1$ & $\tau$ & $q^\nu$ & $\text{e}^{-\nu m}$ & $-c_0\frac{\text{d}\log q}{\text{d}m}$ \\ \hline
GW & $\Tilde{\phi}$ & $\Tilde{\phi}_D$ & $\partial\Tilde{\phi}_D/\partial\Tilde{\phi}$ & $c$ & $q_{\Tilde{\phi}}$ & $C_{\Tilde{\phi}\Tilde{\phi}\Tilde{\phi}}$ \\ \hline
\end{tabular}
\caption{The correspondence between Mahler and GW in the tropical limit.}\label{dictionary}
\end{table}

Regarding the dictionary in Table \ref{dictionary}, we expect that the correspondence between Mahler measure and GW invariants holds for all 16 reflexive polygons. It would be interesting to have a precise proof of the correspondence. Incidentally, the partition function on $S^2$ for certain gauged linear sigma model was used to compute genus-0 GW invariants for a 3d CY variety\footnote{Notice that the CY varieties studied are all compact, though some discussions are made in the large volume regime.} in \cite{Jockers:2012dk} without the use of mirror symmetry. In particular, this linear sigma model flows an IR non-linear sigma model with the CY variety as the target space. It would be interesting to see whether modular Mahler measures could have any relations to this. Moreover, the dictionary between Mahler measure and GW invariants can be potentially extended to the topological vertex formalism.

By virtue of elliptic curves, the theories discussed in this section would have natural connections to Seiberg-Witten (SW) theories as pointed out in \cite{Lerche:1996ni,Hori:2000ck}. It is also worth noting that dessins have also appeared in the study of SW curves as in \cite{Ashok:2006br,He:2015vua,Bao:2021vxt}. It could be possible that the discussions on dessins and (modular) Mahler measures in this paper would give some new insights to the study of SW theories and topological strings.
From the perspective of (modular) Mahler measure, it would also be interesting to apply this to crystal melting, superconformal index, knot/quiver correspondence, black holes etc.

\linespread{0.9}\selectfont
\addcontentsline{toc}{section}{References}
\bibliographystyle{utphys}
\bibliography{ref}

\end{document}